\documentclass{article}
\usepackage{arxiv}
\usepackage[utf8]{inputenc} 
\usepackage[T1]{fontenc}    
\usepackage{hyperref}       
\usepackage{url}            
\usepackage{booktabs}       
\usepackage{amsfonts}       
\usepackage{nicefrac}       
\usepackage{microtype}      
\usepackage{lipsum}
\usepackage{graphicx}
\usepackage[numbers]{natbib}
\usepackage{float}
\graphicspath{ {./images/} }
\title{Fast Finite-Time Sliding Mode Control for Chattering-Free Trajectory Tracking of  Robotic Manipulators }

\author{Momammad Ali Ranjbar,\\ 
  Department of Electrical and Computer Engineering\\
  Tarbiat Modares University\\
  Tehran, Iran \\
  \texttt{ma.ranjbar@modares.ac.ir} \\
}

\usepackage{amsmath}
\begin{document}

\maketitle

\begin{abstract}
Achieving precise and efficient trajectory tracking in robotic arms remains a key challenge due to system uncertainties and chattering effects in conventional sliding mode control (SMC). This paper presents a chattering-free fast terminal sliding mode control (FTSMC) strategy for a three-degree-of-freedom (3-DOF) robotic arm, designed to enhance tracking accuracy and robustness while ensuring finite-time convergence. The control framework is developed using Newton-Euler dynamics, followed by a state-space representation that captures the system's angular position and velocity. By incorporating an improved sliding surface and a Lyapunov-based stability analysis, the proposed FTSMC effectively mitigates chattering while preserving the advantages of SMC, such as fast response and strong disturbance rejection. The controller's performance is rigorously evaluated through comparisons with conventional PD sliding mode control (PDSMC) and terminal sliding mode control (TSMC). Simulation results demonstrate that the proposed approach achieves superior trajectory tracking performance, faster convergence, and enhanced stability compared to existing methods, making it a promising solution for high-precision robotic applications.
\end{abstract}


\section{Introduction}
Today's robots, which incorporate numerous electronic and mechanical components, were initially designed as robotic arms to manipulate objects and perform simple, repetitive tasks that often lead to worker fatigue, loss of concentration, and reduced productivity. These robots, commonly referred to as mechanical arms, are typically installed at fixed locations with predefined operational zones in various industrial settings, such as factories, where they execute tasks like packaging, assembling components, handling heavy objects, and performing high-precision processes such as welding and painting \cite{sadegh2014modeling}. These robots consist of multiple interconnected links joined by several joints, forming a mechanical structure analogous to a human arm. The complexity of a robotic arm's configuration is determined by its intended functions and movement capabilities. Additionally, the number of joints, links, and physical constraints are key factors in defining a robot's degrees of freedom (DOF) \cite{merat1987introduction}.

Doan et al. \cite{doan2020novel} present a fast terminal sliding mode control (FTSMC) scheme for a mechanical arm robot. To accelerate convergence time in the presence of uncertainties and enhance position-tracking accuracy, a fast terminal sliding mode manifold (FTSMM) is developed. A Stimulated Interlocking Control Law (STCL) is incorporated to address the nonlinear and unknown aspects of the system. This technique effectively compensates for external disturbances and uncertain system dynamics in each iteration of the algorithm. A notable advantage of this control scheme is its ability to function without requiring an exact dynamic model of the system, even in the presence of uncertainties. Furthermore, the proposed controller applies smoother control torque commands with reduced oscillations. However, it does not adequately address tracking error, which does not converge to zero.

Liu \cite{liu2019dynamic} proposes a dynamic modeling approach and terminal sliding mode controller (TSMC) for a three-degree-of-freedom mechanical robot to mitigate dynamic errors caused by incomplete modeling due to the system's nonlinearity and multivariable nature. The dynamic model is derived using Lagrange's equations, and a TSMC is designed based on these equations while accounting for uncertainties such as modeling errors and external disturbances. This approach ensures accurate trajectory tracking. Additionally, the stability of the closed-loop system is verified using a Lyapunov function. However, the system's settling time remains an issue that is not effectively addressed.

Van et al. \cite{van2020self} introduce a novel control scheme that integrates derivative-proportional-integral (PID) controllers with a non-singular fast terminal sliding mode (NFTSM) approach and employs time delay estimation (TDE) and fuzzy logic. The resulting control system, termed STF-PID-NFTSMC, exhibits several advantages, including faster transient response with finite-time convergence, minimal steady-state error, and reduced chattering. However, the system's performance is highly dependent on the selection of PID controller coefficients. Additionally, the control design necessitates an accurate dynamic model of the robot. The fuzzy logic mechanism enables a self-adjusting sliding surface, while the TDE algorithm estimates the system's dynamic model without requiring prior knowledge, thereby reducing computational complexity.

Goel et al. \cite{goel2017mimo} propose an adaptive high-order super twisting sliding mode control for uncertain nonlinear systems, providing a robust, continuous control method with finite-time convergence. This approach combines a nonlinear homogeneous sliding surface, conventional super twisting sliding mode control, and an adaptive mechanism to enhance fast convergence, eliminate disturbances, smooth control input, and minimize chattering. The methodology is applied to control two-degree-of-freedom robots.

Rahmani et al. \cite{rahmani2020new} present a novel sliding mode controller (NSMC) for a two-degree-of-freedom robot, leveraging the extended gray wolf optimization (EGWO) algorithm. Since the conventional proportional-derivative (PD) control method is less resistant to external disturbances compared to sliding mode control (SMC), a hybrid approach combining PD and SMC is introduced to compensate for each method's drawbacks. The optimization of control parameters is achieved using GWO and EGWO with weighted coefficients. The stability of NSMC is demonstrated via a Lyapunov function, and its performance is compared against conventional SMC and Proportional-Derivative Sliding Mode Control (PDSMC).

Medjebouri et al. \cite{medjebouri2016adaptive} successfully employ an adaptive neural network controller based on SMC for controlling the PUMA mobile robot's path tracking. The neural structure learns nonlinear system dynamics through layered feature extraction and feedback mechanisms similar to those described in the \cite{norouzi2025primer}, where feedforward and recurrent networks are shown to approximate complex input–output mappings \cite{alizadeh2025epidemic}. Simulation results indicate that the adaptive sliding mode controller effectively reduces chattering and improves path tracking performance compared to conventional SMC. Additionally, the number of required control inputs is reduced, thereby enhancing energy efficiency.

Truong et al. \cite{truong2019adaptive} propose an adaptive fuzzy sliding mode controller for a three-degree-of-freedom hydraulic robotic arm under significant load variations. A disturbance observer is designed to detect changes in the mass of external cargo, utilizing electrohydraulic actuators as primary torque generators. The proposed control scheme integrates a backward sliding mode controller, fuzzy logic system (FLS), and a nonlinear disturbance observer. The FLS adjusts the control gain based on the observer's output to compensate for load variations. The stability of the system is established using a Lyapunov function.

Amer et al. \cite{amer2011adaptive} discuss the control of a three-degree-of-freedom industrial robot with nonlinear dynamics in the presence of uncertainties and external disturbances. A control strategy combining a fuzzy logic controller and SMC is proposed to mitigate chattering effects. The sliding-fuzzy-adaptive mode control (AFSMC) incorporates a derivative-proportional-integral controller to determine the sliding surface. Stability is guaranteed using the Lyapunov stability theorem, ensuring that all signals remain bounded and the system output asymptotically tracks the desired reference trajectory.

Spong \cite{spong2020robot} presents a comprehensive reference on robotic dynamics, including forward and inverse kinematics, velocity kinematics, the Jacobian matrix, path planning, and robot arm trajectory tracking.
Furthermore, there are also have been a lot of work leveraging learning methods to extract data \cite{ sholehrasa2024integratingproteinsequenceexpression}, motion planning, and control of the robots, however the performance of these methods are not promised in all situations and highly depends on the input data, hyperparameters tuning and training methods, which might not be the best solution for controlling robotic arms, which requires a promised and reliable performance to both guarantee safety of the workforce\cite{ correll2022introduction}.

This paper explores various control methodologies for mechanical arms, each tailored to specific system requirements and objectives. The focus is on designing a fast terminal sliding mode controller (FTSMC) for controlling the rotational angles of a 3-DOF robotic arm. SMC is widely utilized due to its robustness and simplicity. The primary goal is to achieve system asymptotic stability through conductivity control and equivalent control components. However, conventional asymptotic stability may not ensure rapid convergence, necessitating resistive control forces. Nonlinear control techniques, such as terminal sliding mode control, significantly enhance transient performance. Compared to classical SMC, finite-time convergence provides improved performance when initial states are far from the set-point. Thus, this study proposes an FTSMC that integrates the benefits of both TSMC and classical SMC, ensuring rapid finite-time convergence under varying conditions.

The remainder of this article is structured as follows: Section 2 discusses the dynamic equations of a 3-DOF robot. Section 3 introduces system modeling and derives the state-space equations. Sections 4 and 5  present the fundamentals of FTSMC and stability proof, followed by control input design. Section 6 details software simulations under different conditions. Finally, Section 7 provides a summary and conclusion.

\section{Robotic arm dynamic model}\label{sec:dynamic_equations}
A planar rotational 3-DOF robot consists of three rotating joints and three links, as shown in figure 1; at the end of the third link, an End Effector is installed to carry loads. Also, the figure indicates that this robot is connected to a base at joint-1. According to \cite{doi:10.1177/09596518241229741} proposed model for the 3-DOF robotic arm, the dynamic differential equation model captures the complex motion and interaction of forces in robotic arms. This equation is crucial for understanding and controlling robotic arm dynamics, as it comprehensively describes the arm's motion by accounting for three critical dynamic components:
\begin{equation} 
M(\theta)\ddot{\theta} + C(\theta,\Dot{\theta}) + G(\theta) = \tau
\end{equation}
Where \( \theta = (\theta_1, \theta_2, \theta_3 )^T \in \mathbb{R}^{3 \times 1} \) corresponds to the angular positions for the arms, 
\( \dot{\theta} = (\dot{\theta}_1, \dot{\theta}_2, \dot{\theta}_3 )^T \in \mathbb{R}^{3 \times 1} \) corresponds to the angular velocity vector of order \( 3 \times 1 \), 
and \( \ddot{\theta} = (\ddot{\theta}_1, \ddot{\theta}_2, \ddot{\theta}_3 )^T \) represents the angular acceleration vector.

\( M(\theta) \in \mathbb{R}^{3 \times 3} \) is the actual inertia matrix of order \( 3 \times 3 \), 
\( C(\theta, \dot{\theta}) \in \mathbb{R}^{3 \times 1} \) is the \( 3 \times 1 \) inertia matrix representing centrifugal and Coriolis forces, 
and \( G(\theta) \in \mathbb{R}^{3 \times 1} \) is the inertia vector of order \( 3 \times 1 \), which represents the gravity matrix.

Finally, \( \tau \) is the system control input.
This paper aims to implement accurate and fast trajectory tracking of robotic arms with minimum error.
Lemma 1, For the following nonlinear system \cite{angulo2013robust}:
\begin{equation} 
\dot{x}=f(t,x), x(0)=x_0, x\in R^n
\end{equation}
Where \( f(t,x) \) is a continuous function. If \( V(x) \) is considered as a Lyapunov function which satisfies:
1. \( V(x) = 0 \leftrightarrow x = 0 \),
2. \( \dot{V}(x) \leq a V^{\gamma_1}(x) - b V^{\gamma_2}(x) \),

where \( a, b > 0 \), \( \gamma_1 > 1 \), and \( 0 < \gamma_2 < 1 \). 

Then, the system defined in equation (2) is finite-time stable. Moreover, if \( T \) represents the settling time of the system, it will satisfy:
\[
T \leq \frac{1}{a(\gamma_1 - 1)} + \frac{1}{b(1 - \gamma_2)}.
\]
According to Lemma 1, if there is a Lyapunov function V(x), the finite time convergence of FSTMC is promised.
\begin{figure}
	\centering
	\includegraphics[width=0.8\linewidth]{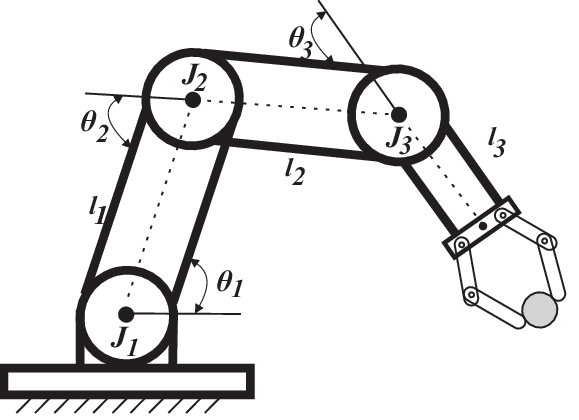}
        	\caption{Scheme of 3-DOF arm robot \cite{doi:10.1177/09596518241229741} }
	\label{}
\end{figure}

\section{State-space equations} \label{{sec:modeling}}
First, by rewriting the dynamic equations governing 3-DOF robotic arms in equation (1) as follows:
\begin{equation} 
M(\theta)\ddot{\theta} + B(\theta)[\dot{\theta}\dot{\theta}] + D(\theta)[\dot{\theta}^2] + G(\theta) = \tau
\end{equation}
Where $\ddot{\theta}$, [$\dot{\theta}\dot{\theta}$], and [$\dot{\theta}^2$] are considered as equation (3). furthermore, based on \cite{merat1987introduction}, \cite{alizadeh2022fault}, \cite{xu2021nonlinear} the value of the matrices $M(\theta), B(\theta), D(\theta)$, and $G(\theta)$, which depend on the robot parameters, are calculated and would be utilized in the simulations using the values presented in table I.
\begin{equation} 
 \begin{cases}
 \ddot{\theta} = \bigg[ \ddot{\theta}_1 \quad  \ddot{\theta}_2 \quad  \ddot{\theta}_3 \bigg]^T \\
[\dot{\theta}\dot{\theta}] = \bigg[ \dot{\theta}_1\dot{\theta}_2 \quad  \dot{\theta}_1\dot{\theta}_3 \quad  \dot{\theta}_2\dot{\theta}_3 \bigg]^T \\
[\dot{\theta}^2] = \bigg[ \dot{\theta}_1^2 \quad  \dot{\theta}_2^2 \quad  \dot{\theta}_3^2 \bigg]^T 
 \end{cases}
\end{equation}
By considering the following equation:
\begin{equation} 
 \begin{cases}
X = [ x_1 \quad  x_2 \quad  x_3 \quad x_4 \quad  x_5 \quad  x_6 ]^T =\\
[ \theta_1 \quad  \dot{\theta}_1 \quad  \theta_2 \quad \dot{\theta}_2 \quad  \theta_3 \quad  \dot{\theta}_3 ]^T \\
\tau = [\tau_1 \quad  \tau_2 \quad  \tau_3]^T = [u_1 \quad  u_2 \quad  u_3]^T
 \end{cases}
\end{equation}
Where $X \in R^{6 \times 1}$ is the state space vector and $\tau \in R^{3 \times 1}$ is the torque vector linked to the control inputs. To facilitate the development of the control method, equation (3) can be rewritten in the following form \cite{mien2014adaptive}:
\begin{equation} 
 \begin{cases}
 \dot{x_1} = \dot{x_2}\\
 \dot{x_2} = f_1(x) + g_1(x)u_1\\
 \dot{x_3} = x_4\\
 \dot{x_4} = f_2(x) + g_2 (x)u_2\\
 \dot{x_5} = x_6\\
 \dot{x_6} = f_3(x) + g_3(x)u_3
 \end{cases}
\end{equation}

so that $f_1 (x),f_2 (x),f_3 (x),g_1 (x),g_2 (x)$, and $g_3 (x)$ are defined as follows:
\begin{align}
f_i(x) &= -\big[M^{-1}(\theta) (B(\theta)[\dot{\theta} \dot{\theta}] + C(\theta)[\dot{\theta}^2] + G(\theta))\big]_i, \nonumber \\
i &= 1, 2, 3 \nonumber \\
g_i(x) &= \big[M^{-1}(\theta)\big]_i, \quad i=1, 2, 3
\end{align}

\begin{table}
\caption{Considered parameters for the simulation and controller design.\label{T1}}
\begin{tabular}{ll}
\toprule
\midrule
parameters&$m_1=m_2=m_3=1kg$,\\
&$l_1=0.5m, l_2=l_3=1m$\\
&$g=9.81m/s^2$\\
optimum state of&$\theta_{1d}=0.35-0.5sint,$\\
robot's angles&$\theta_{2d}=0.25+0.5cost,$\\
&$\theta_{3d}=0.45-0.5cost$\\
initial states of&$\theta_1(0)=0.7, \theta_2(0)=1.5,$\\
robot's angles&$\theta_3(0)=0.5$\\
\bottomrule
\end{tabular}
\end{table}
The mass matrix of the mechanical arm $M(\theta)$, consists of all terms in the class of equations (3) that are multiples of $\ddot{\theta}$ The gravitational term, $G(\theta)$, includes all the terms in equation (3) in which $g$, the earth's gravitational constant, appears. Notably, the gravitational term depends only on $\theta$, not its derivatives. $B(\theta)$ is a 3×3 matrix of Coriolis coefficients, and $C(\theta)$ is a 3x3 matrix of centrifugal coefficients.

\section{Fast terminal sliding mode control} \label{{sec:FTSMC}}

In this part, a conventional terminal sliding mode control (TSMC) for the first arm is proposed for the first arm of the robot. The proposed sliding surfing is as: 
\begin{equation} 
S = \dot{e}_{1} + \beta(e_{1})^{\frac{p}{q}}
\end{equation}
\\
Where $e_{1}=x_{1}-x_{1d}$ and $x_{1d}$ are desired angular positions of the first arm, shown in table I. Similarly, $\beta > 0$ also parameters $p$ and $q$ are integers where $q$$<$$p$. Nonlinear term $(e_{1} )^{\frac{p}{q}}$ guarantees convergence in finite time which can be calculated using lemma 1, as $t_s=(\frac{p}{((\beta(p-q))}\big|e_1(0)\big|^\frac{(p-q)}{p}$, to the coordinate origin; Therefore, the convergence time can be adjusted by modifying  $\beta, q, $and $p$. The drawback of the mentioned sliding surface is that the convergence time depends on the initial distance from the sliding surface; Consequently, as the distance increases, the convergence time will increase proportionally to $\big|e_1(0)\big|^\frac{(p-q)}{p}$. Therefore, a fast terminal SMC (FTSMC) is proposed to solve this problem. The designed sliding surface for FTSMC is as follows:

\begin{equation} 
S_1 = \dot{e}_{1} + \alpha(e_{1}) + \beta(e_{1})^{\frac{p}{q}}
\end{equation}

Where $\alpha$ is a positive constant ($\alpha>0$), next, the If the system is on a sliding surface, then $\dot{e}_{1} = \alpha e_{1} - \beta(e_{1})^{\frac{p}{q}}$) is established. Now, it will be discussed how the proposed sliding surface will reduce convergence time. If the initial state of the system is far from the origin, the approximate dynamics is $\dot{e}_{1} = \alpha e_{1}$, and if the initial state is close to the origin, then the approximate dynamics is $\dot{e}_{1} = -\beta(e_{1})^{\frac{p}{q}}$). Using parameters $\alpha$ and $\beta$, the convergence time can be set independently, whether far or close to the origin. It can be shown that the convergence time is obtained from the following equation (10)  \cite{yu2002fast}

\begin{equation} 
t_s = \frac{p}{\alpha(p-q)}(ln(\alpha \quad e_1(0)^{\frac{(p-q)}{p}} + \beta) - ln\beta)
\end{equation}

\section{Control law design}
Due to the similarity of the controller designing principle for all three arms, control designing is only discussed for the first arm’s angular position, and the repetition of similar content is avoided.

and the deviation of equation (9), the following equation is obtained:
\begin{equation} 
\dot{S}_1(x) = \ddot{e}_{1} + \dot{e}_{1}(\alpha+\beta\frac{p}{q}(e_{1})^{\frac{p}{q}-1})
\end{equation}
Assuming that the sliding surface equation is defined based on state variables and desired state, the system converges to the desired shape on the sliding surface. So, it can be said that the following relations are established on the sliding surface:
\begin{equation} 
 \begin{cases}
S_1 = 0\\
 \dot{S}_1 = 0
 \end{cases}
\end{equation}
According to the sliding mode theory, the control input $u$ is defined as follows \cite{qiao2007indirect}, \cite{spong2020robot}, \cite{zhai2021fast}:
\begin{equation} 
u = u_{eq} + u_s
\end{equation}
 So that $u_{eq}$ is the equivalent control component and maintains the states on the sliding surface, and $u_s$ is the switching and directs the states to the sliding surface, which $u_s$ is responsible for stabilizing the system and is determined by the Lyapunov stability criterion. We adopted the sliding mode control methodology discussed by Samaei et al.  \cite{samaei2023comment}, which provides an approach to ensuring both asymptotic convergence to the sliding surface and robust maintenance of the system's trajectory within that surface.
 By setting $\dot{s_1}=0$, the equivalent control law is obtained as follows:
\begin{align}
\dot{S}_1(x) &= \ddot{e}_1 + \dot{e}_1 \left(\alpha + \beta \frac{p}{q} e_1^{\frac{p}{q}-1} \right) = 0  \Rightarrow \nonumber \\
f_1(x) &+ g_1(x) u_{1eq} + \dot{e}_1 \left(\alpha + \beta \frac{p}{q} e_1^{\frac{p}{q}-1} \right) = 0 \nonumber \\
\Rightarrow & \quad u_{1eq} = \frac{-f_1(x) - \dot{e}_1 \left(\alpha + \beta \frac{p}{q} e_1^{\frac{p}{q}-1} \right)}{g_1(x)}
\end{align}

Thus, a suitable candidate for the switching control law is shown in the following equation:
\begin{equation} 
u_{1s} = -k_1sgn(S_1)
\end{equation}
where $k_1$ is a positive real value. By combining equations (14), (15), (18)-(21) and inserting them into equation (13), the general law of fast terminal sliding mode robust control, which is applied to the first arm of the robot to follow the desired reference angle, is:
\begin{equation} 
u_1 = \frac{-f_1(x)-\dot{e}_{1x}(\alpha+\beta\frac{p}{q}(e_{1x})^{\frac{p}{q}-1})}{g_1(x)} - k_1sgn(S_1)
\end{equation}

The aim of this paper is to obtain zero trajectory tracking error within a finite reaching time to the sliding surface. Before moving any further, the following assumption is imposed.\\

**Assumption 1:** The matrix \( M(\theta) \) and \( C(\theta, \dot{\theta}) \), along with their derivatives, are bounded.\\

**Lemma 2:** For any \( x \in \mathbb{R}, i = 1,2,\dots,n \) and \( p \) as a positive constant \cite{zhai2021fast}, 
\begin{equation} 
(|x_1| + \dots + |x_n|)^p < \max\{n^{p-1},1\} (|x_1|^p + \dots + |x_n|^p).
\end{equation}

Theorem 1, For the system equation (1), by utilizing equation (9) as the sliding surface, the FTSMC control law is proposed in equation (16), then the system will reach the designed sliding surface in a finite time $t_s$, and the trajectory tracking error of the sliding surface will be 0 in finite time $t_s$\cite{samaei2023comment}.\\
Proof of Theorem 1: The stability analysis of the FSTM can be discussed as follows:\\
By selecting a Lyapunov function as: 
\begin{equation} 
V_1 = \frac{1}{2}s_1^2 > 0
\end{equation}
The derivation of the proposed Lyapunov function is as follows:
\begin{align}
\dot{V}_1 = S_1 \dot{S}_1 = &S_1 \left( \ddot{e}_1 + \dot{e}_1 \left( \alpha + \beta \frac{p}{q} \left( e_1 \right)^{\frac{p}{q}-1} \right) \right) = \nonumber \\
& S_1 \left( f_1(x) + g_1(x) \left( u_{1eq} + u_{1s} \right) \right) \nonumber \\
& + \dot{e}_1 \left( \alpha + \beta \frac{p}{q} \left( e_1 \right)^{\frac{p}{q}-1} \right) < 0 \\
\notag
\end{align}

Then by using equation (19), and the sign function as a switching control law, and substituting in equation (16):

\begin{align}
\dot{V}_1 = S_1 \dot{S}_1 = &S_1 \left( f_1(x) + g_1(x) \right) \nonumber \\
& \left( \frac{-f_1(x) - \dot{e}_1 \left( \alpha - \beta \frac{p}{q} \left( e_1 \right)^{\frac{p}{q}-1} \right)}{g_1(x)} - k_1 \, \text{sign}(S_1) \right) = \nonumber \\
& \left( \ddot{e}_1 + \alpha(e_1) + \beta e_1^{\frac{p}{q}} \right) \left( f_1(x) + g_1(x) \right) \nonumber \\
& \left( \frac{-f_1(x) - \dot{e}_1 \left( \alpha - \beta \frac{p}{q} \left( e_1 \right)^{\frac{p}{q}-1} \right)}{g_1(x)} - k_1 \, \text{sign}(S_1) \right) \\
\notag
\end{align}

\begin{figure}
	\centering
	\includegraphics[width=1\linewidth]{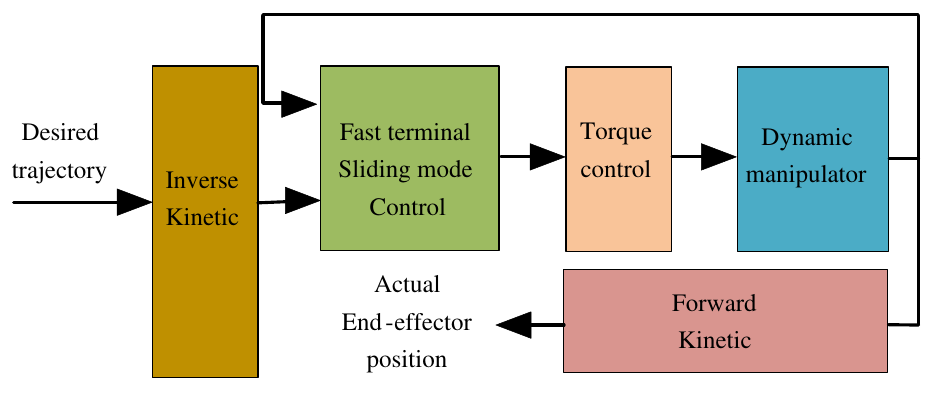}
	\caption{Schematic diagram of the close loop system using FTSMC approach}
	\label{}
\end{figure}
\begin{figure}
	\centering
	\includegraphics[width=1\linewidth]{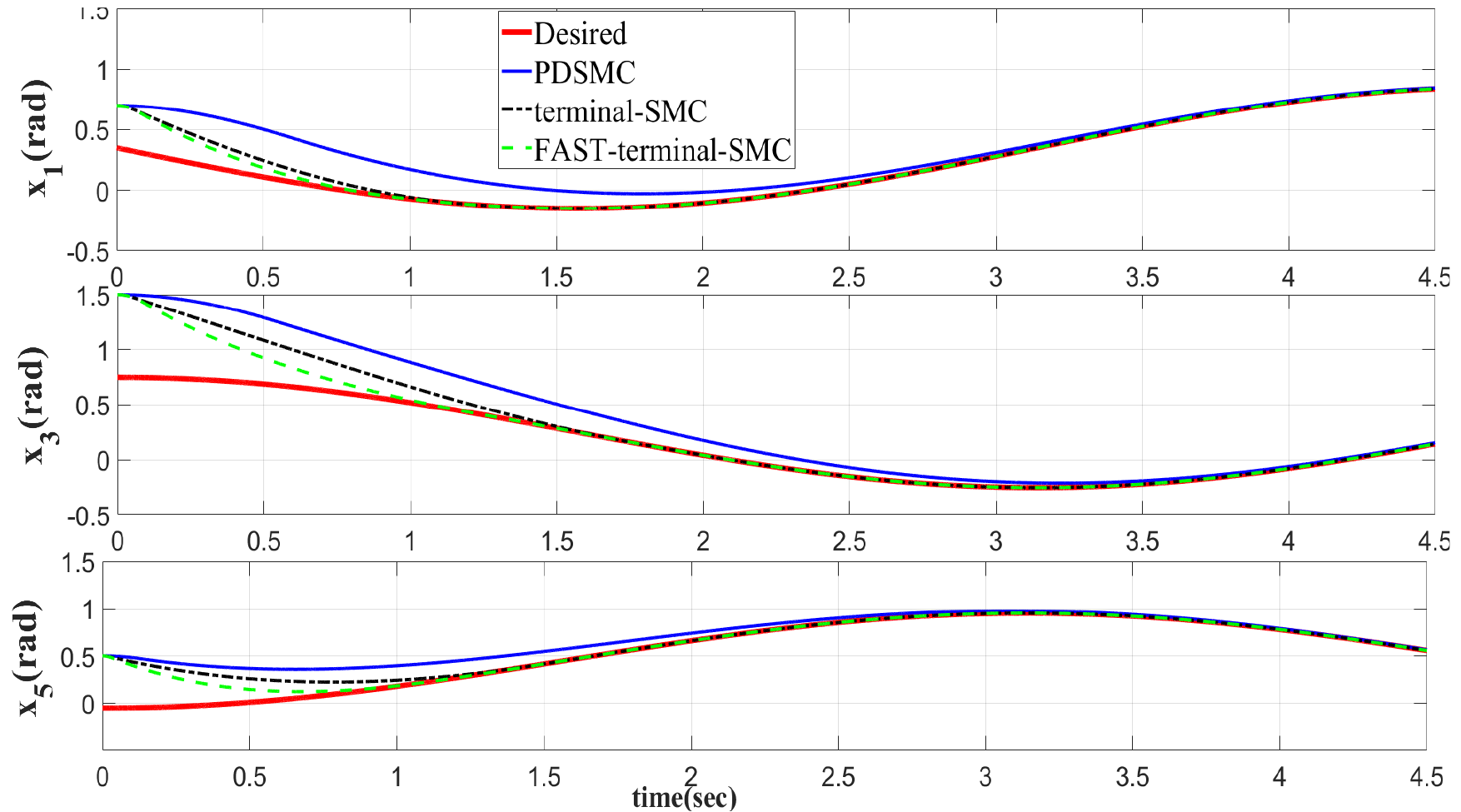}
	\caption{Tracking of robot arms’ angle tracking with conventional PDSMC, the terminal SMC, and the fast terminal SMC with sign function}
	\label{}
\end{figure}
\begin{figure}
	\centering
	\includegraphics[width=1\linewidth]{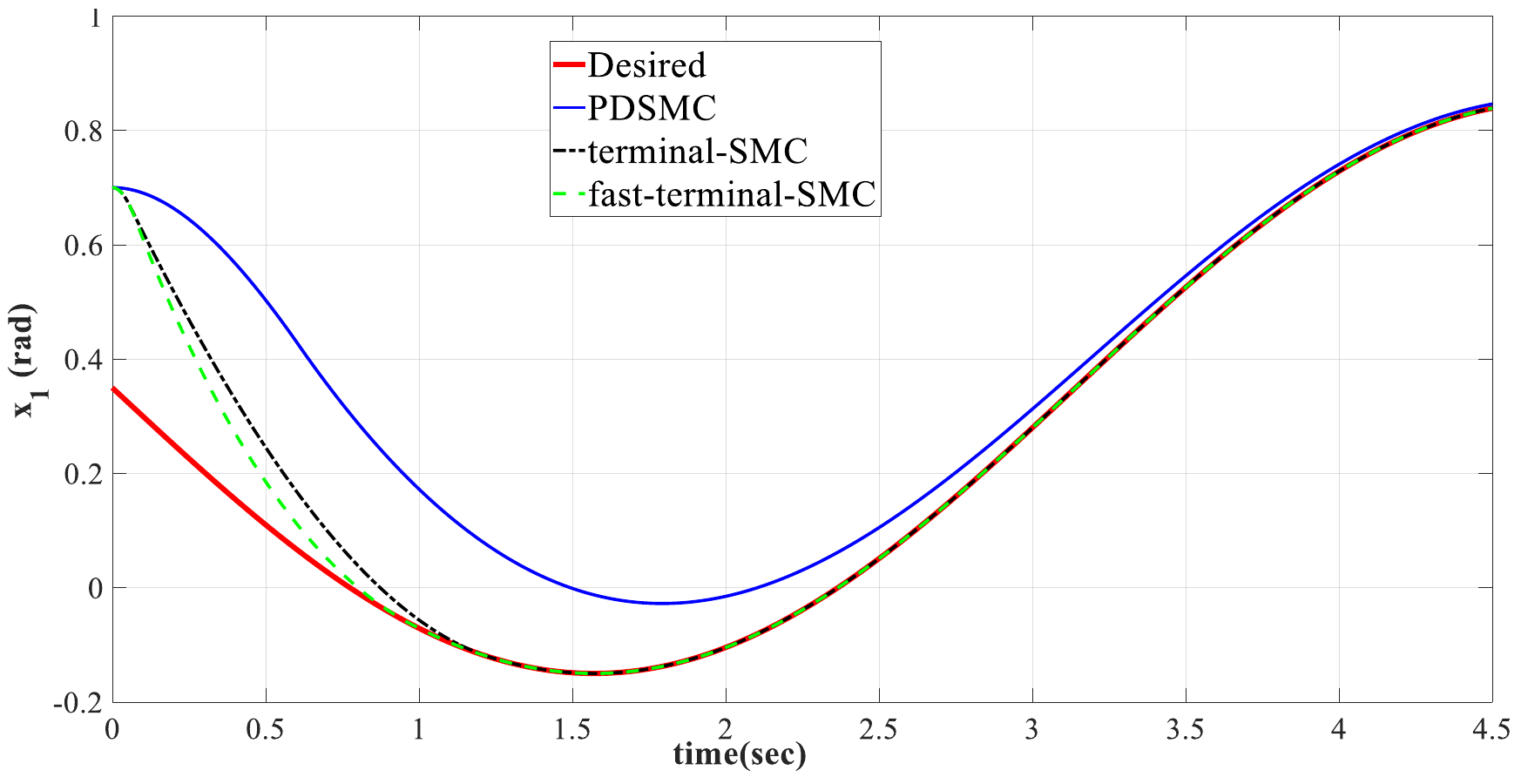}
	\caption{Tracking of the first arm of the robot with conventional PD SMC, the terminal SMC, and the fast terminal SMC with sign function}
	\label{}
\end{figure}
With the help of Lemma 2 and 3 and calculations done by \cite{R15}, it will be as:
\begin{equation} 
\dot{V}_1 \le -\beta\big|\big|e\big|\big|^2 - \alpha^{1-\frac{p}{q}}\big|\big|e\big|\big|^{1-\frac{p}{q}} \le 0
\end{equation}
Where $\dot{V}_1$ can be zero when $e=0$. In conclusion, the trajectory tracking error will quickly converge to zero. Also, it can be helpful to mention that $\alpha$ and $\beta$ are positive, and according to lemma 1, the convergence speed is affected by $\alpha$ and $\beta$, and the larger they are, the faster the convergence speed will be.
Theorem 1 proved this, and the FTSMC approach for robotic arms is summarised in Figure 2. 

\section{Simulation result} \label{{sec:simulation}}
\subsection{Fast terminal sliding mode control with sign function}
Considering the simulation time $T_{sim}=4.5s$ and sampling time  $T_s=5e-4$, and parameter value mentioned in Table 1, and with respect to the control inputs designed discussed in section 5, the following results are obtained.

Figures 3 and 4 show that the trajectory tracking of the robot arms with the fast terminal sliding mode control is achieved in a shorter time and with higher accuracy compared with the conventional PD-sliding mode controller and the terminal sliding mode control. Furthermore, due to greater transparency and similarity of working steps, the following diagrams are given only for the first arm of the robot.
Using a sign function as a switching control law causes high-frequency oscillation known as chattering, which is shown in figure 5.

\begin{figure}
	\centering
	\includegraphics[width=1\linewidth]{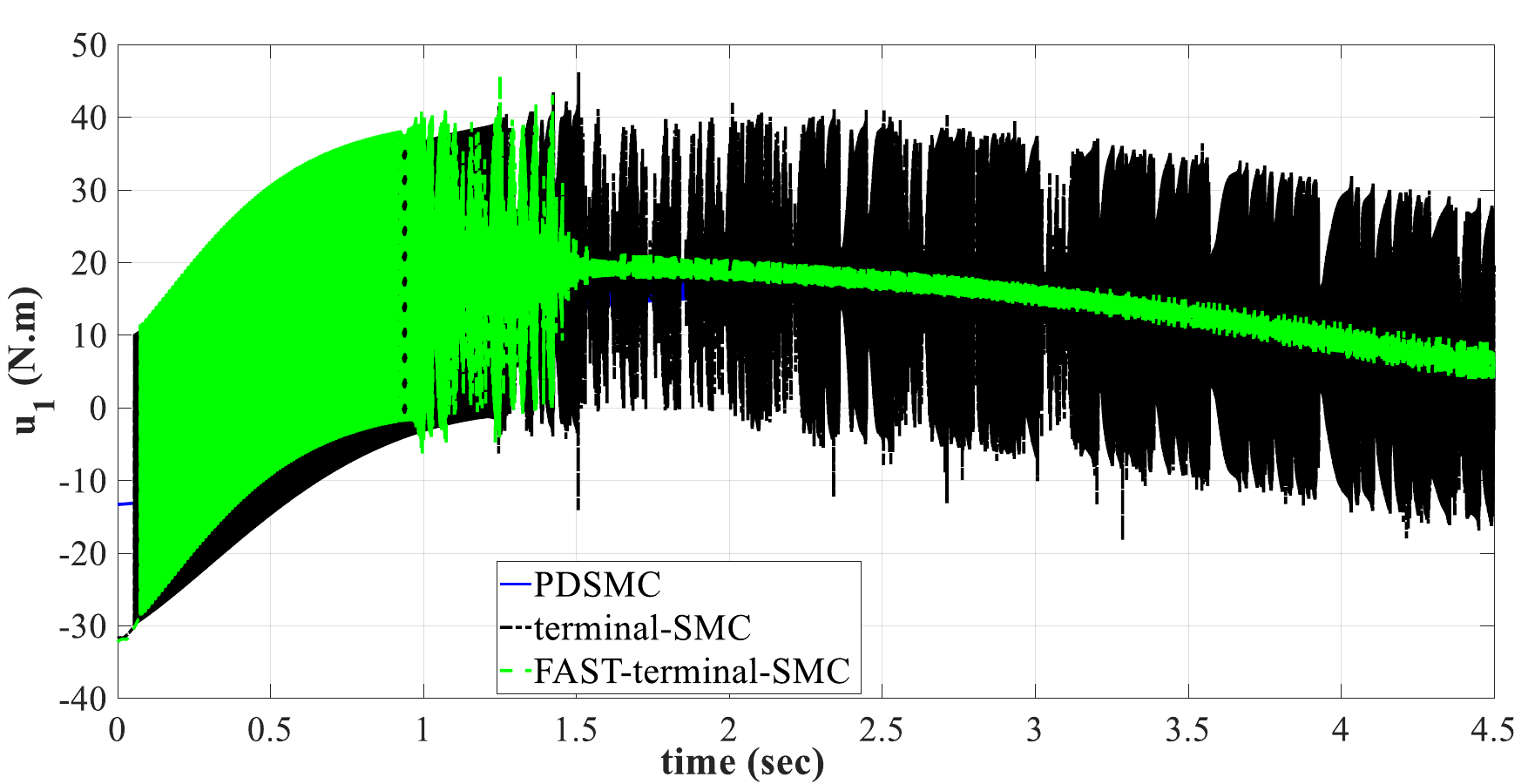}
	\caption{Control effort with conventional PD SMC, the terminal SMC, and the fast terminal SMC with sign function}
	\label{}
\end{figure}
\begin{figure}
	\centering
	\includegraphics[width=1\linewidth]{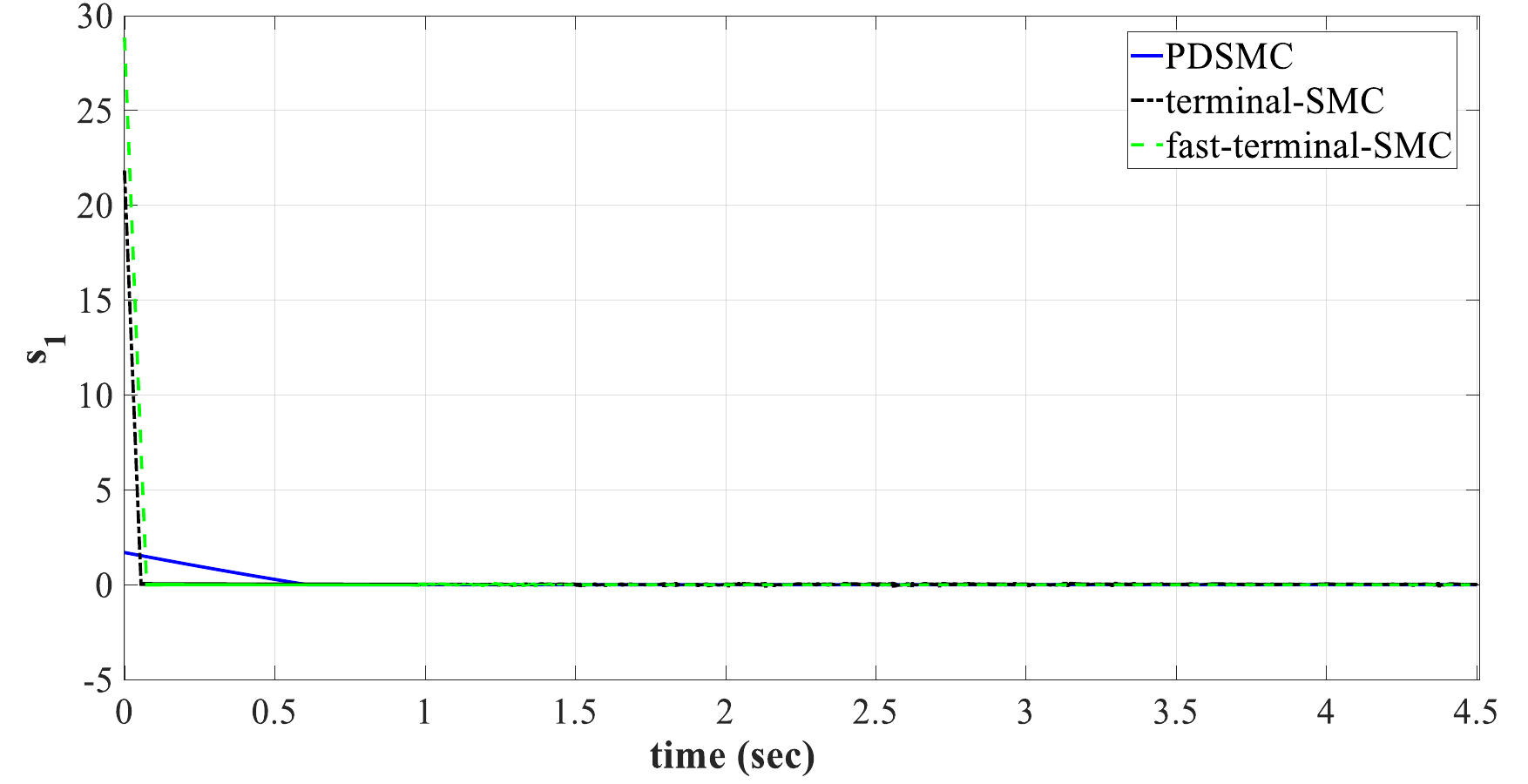}
	\caption{Reaching time to sliding surface with conventional PD SMC, the terminal SMC, and the fast terminal SMC with sign function}
	\label{}
\end{figure}
\begin{figure}
	\centering
	\includegraphics[width=1\linewidth]{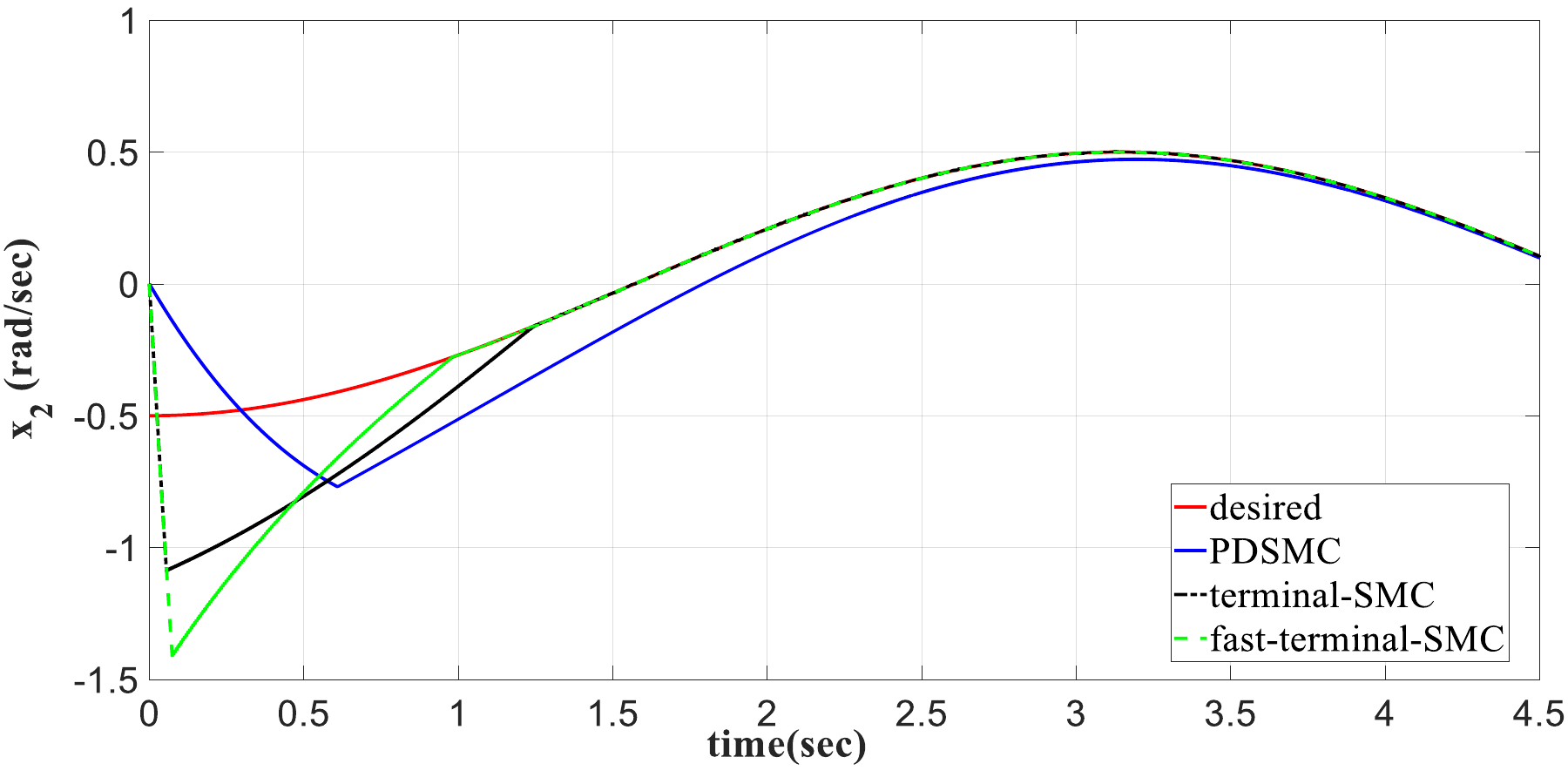}
	\caption{Angular velocity with conventional PD SMC, the terminal SMC, and the fast terminal SMC with sign function}
	\label{}
\end{figure}
\begin{figure}
	\centering
	\includegraphics[width=1\linewidth]{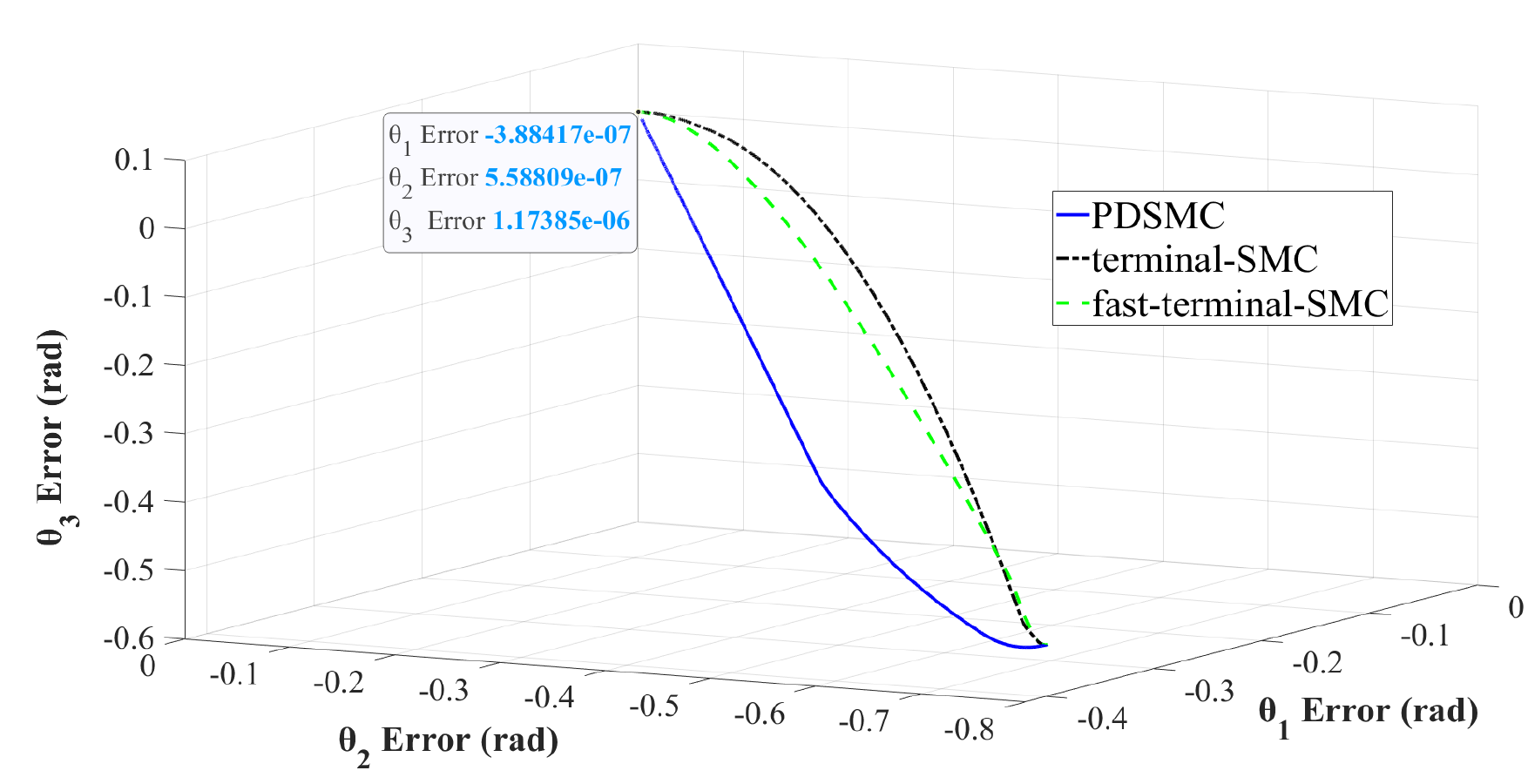}
	\caption{Position tracking error with conventional PD SMC, the terminal SMC, and the fast terminal SMC with sign function}
	\label{}
\end{figure}
\begin{figure}
	\centering
	\includegraphics[width=1\linewidth]{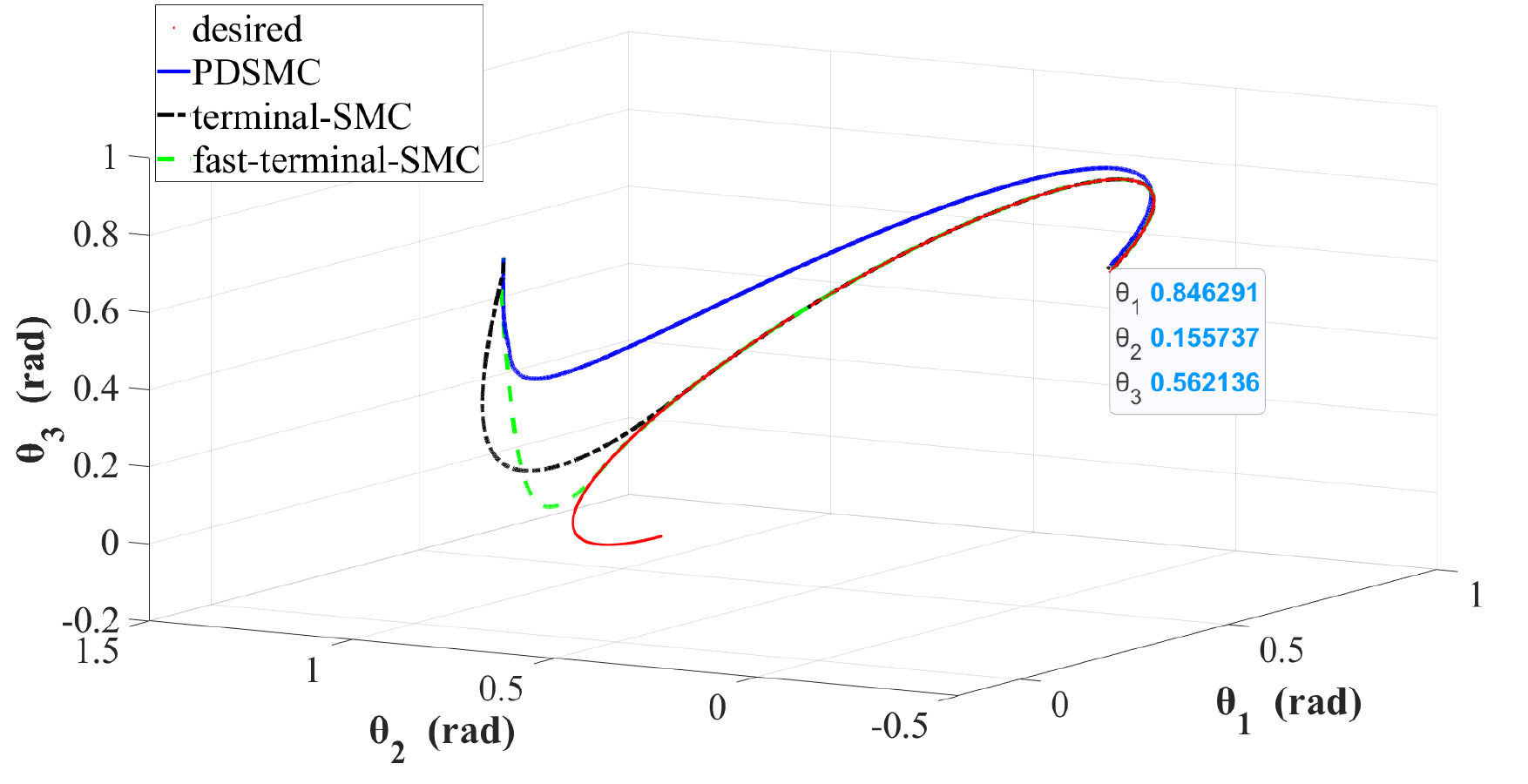}
	\caption{Position tracking with conventional PD SMC, the terminal SMC, and the fast terminal SMC with sign function}
	\label{}
\end{figure}

Figure 6 shows the sliding surface of the robot arm with FTSMC, TSMC, and Conventional PDSMC. The FTSMC reaches the SF in a shorter time than the other two controllers. Angular velocity of the arm again shows that FTSMC has better performance than other controllers. However, it can be seen that there is a rapid change in magnitude in all three controllers, which is mainly caused by using the sign function in the control law.
As it was discussed, the use of sign function causes chattering and sharp changes, particularly in angular velocity, as shown figure 7. The trajectory tracking error of the robot converges to zero, as it is shown in figures 8 and 9. An alternative for the sign function could be used to deal with the chattering problem and smoothen the output. Hyperbolic tangent is similar to sign function, but it is smoother and has no rapid changes, making it an excellent candidate to substitute with sign function.

\begin{table}
\caption{The mean square error values of controllers.\label{T1}}
\begin{tabular}{ll}
\toprule
Control law&MSE\\
\midrule
Fast Terminal Sliding Mode&0.1331\\
Terminal Sliding Mode&0.1976\\
PD-Sliding Mode&0.4698\\
\bottomrule
\end{tabular}
\end{table}

\subsection{Fast terminal sliding mode control with hyperbolic tangent function}
Considering the simulation time of $T_{sim}=4.5s$ and the constant sampling time of $T_s=5e-4$, the parameter values according to table 1, and using hyperbolic tangent as switching control law the following results are obtained.
The trajectory tracking is achieved with high accuracy and in a short time, furthermore, the chattering issue is completely gone, as it is shown in Figures 10 and 11.
Figures 12-15 show that trajectory tracking error converges to zero and the angular velocity is much smoother. FTSMC has better performance in reaching time and trajectory than TSMC and conventional PDSMC, as it was expected.

\begin{figure}
	\centering
	\includegraphics[width=1\linewidth]{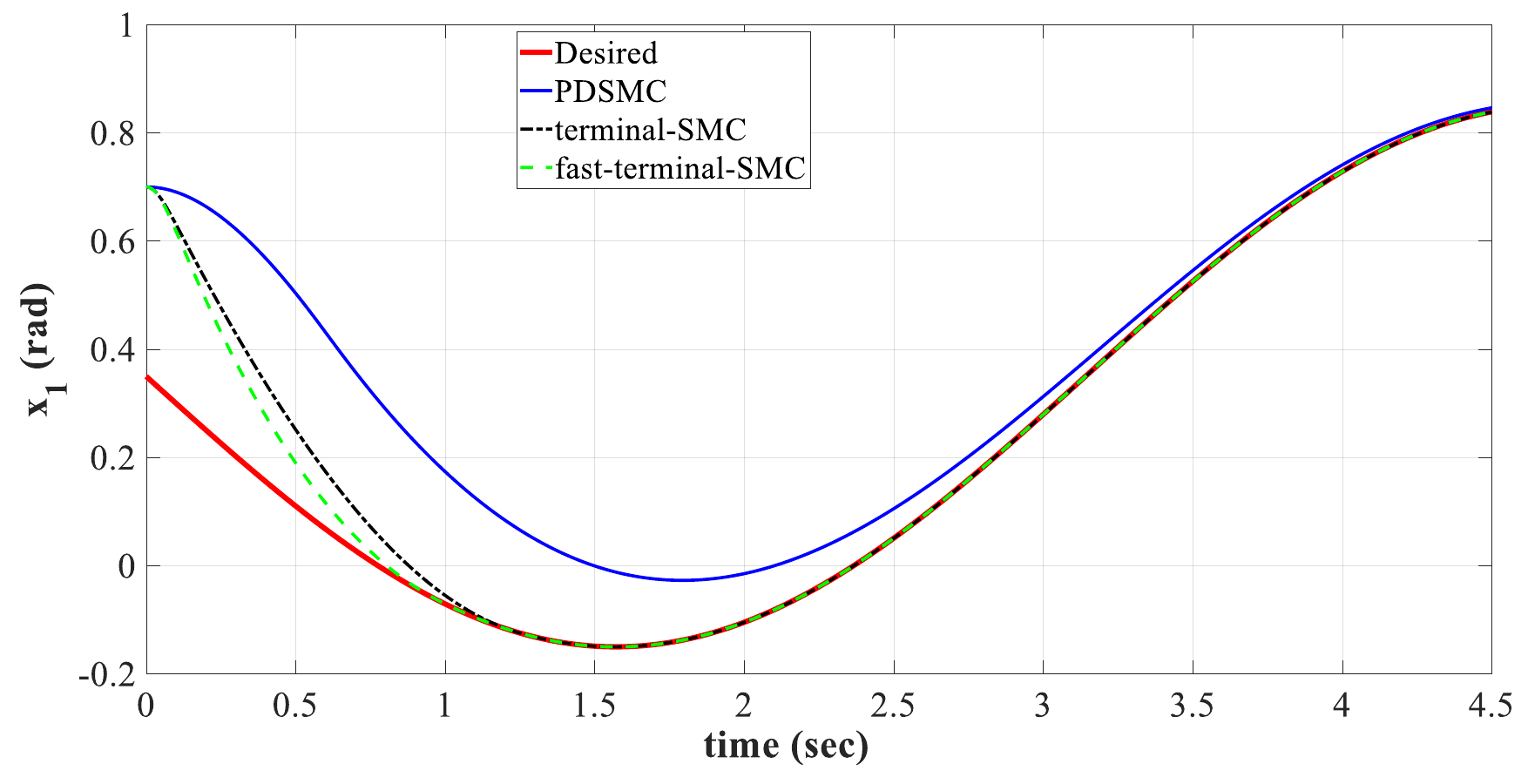}
	\caption{Tracking of the first arm of the robot with conventional PD SMC, the terminal SMC, and the fast terminal SMC with hyperbolic tangent function}
	\label{}
\end{figure}
\begin{figure}
	\centering
	\includegraphics[width=1\linewidth]{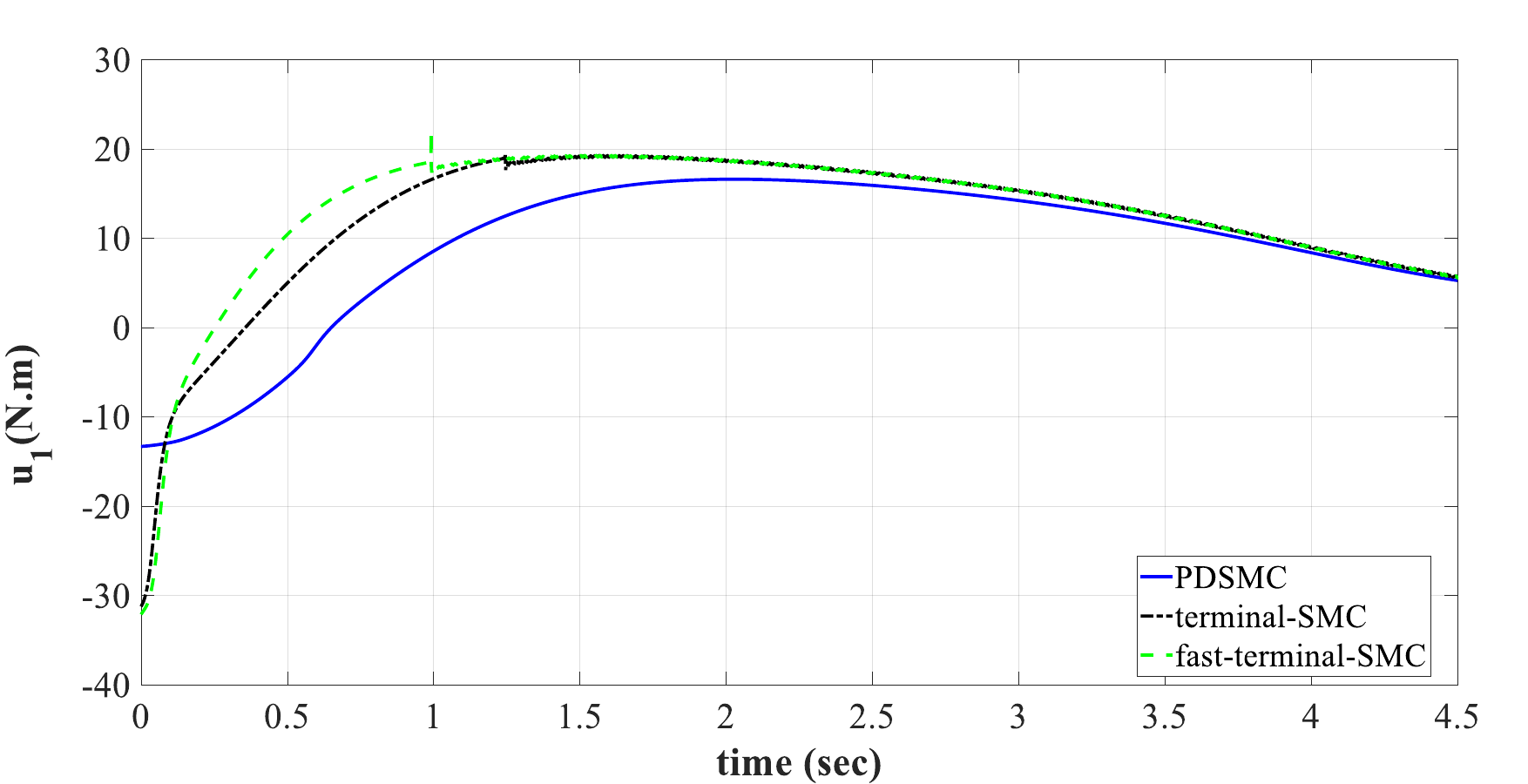}
	\caption{Control effort with conventional PD SMC, the terminal SMC, and the fast terminal SMC with hyperbolic tangent function}
	\label{}
\end{figure}
\begin{figure}
	\centering
	\includegraphics[width=1\linewidth]{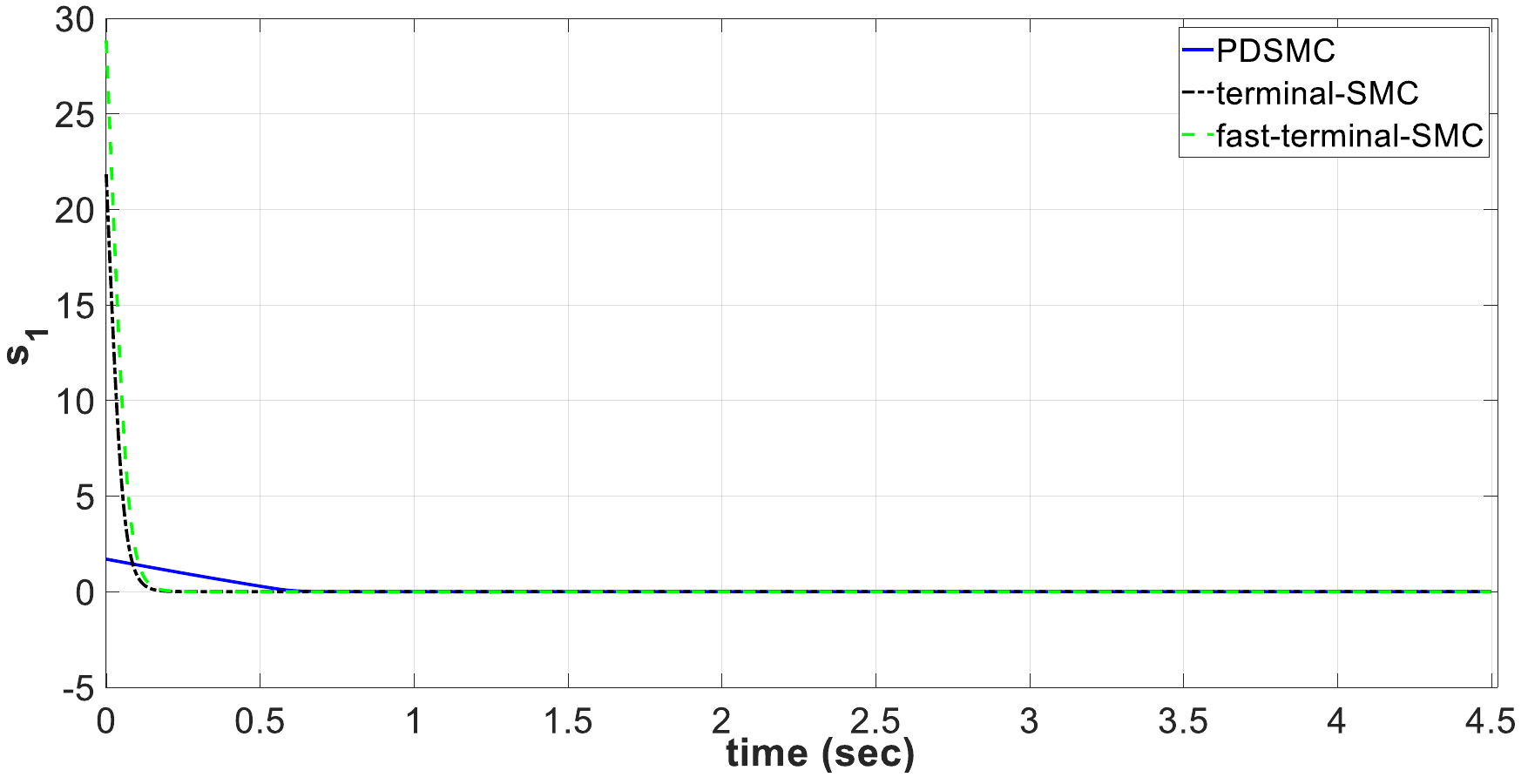}
	\caption{Reaching time to sliding surface with conventional PD SMC, the terminal SMC, and the fast terminal SMC with hyperbolic tangent function}
	\label{}
\end{figure}
\begin{figure}
	\centering
	\includegraphics[width=1\linewidth]{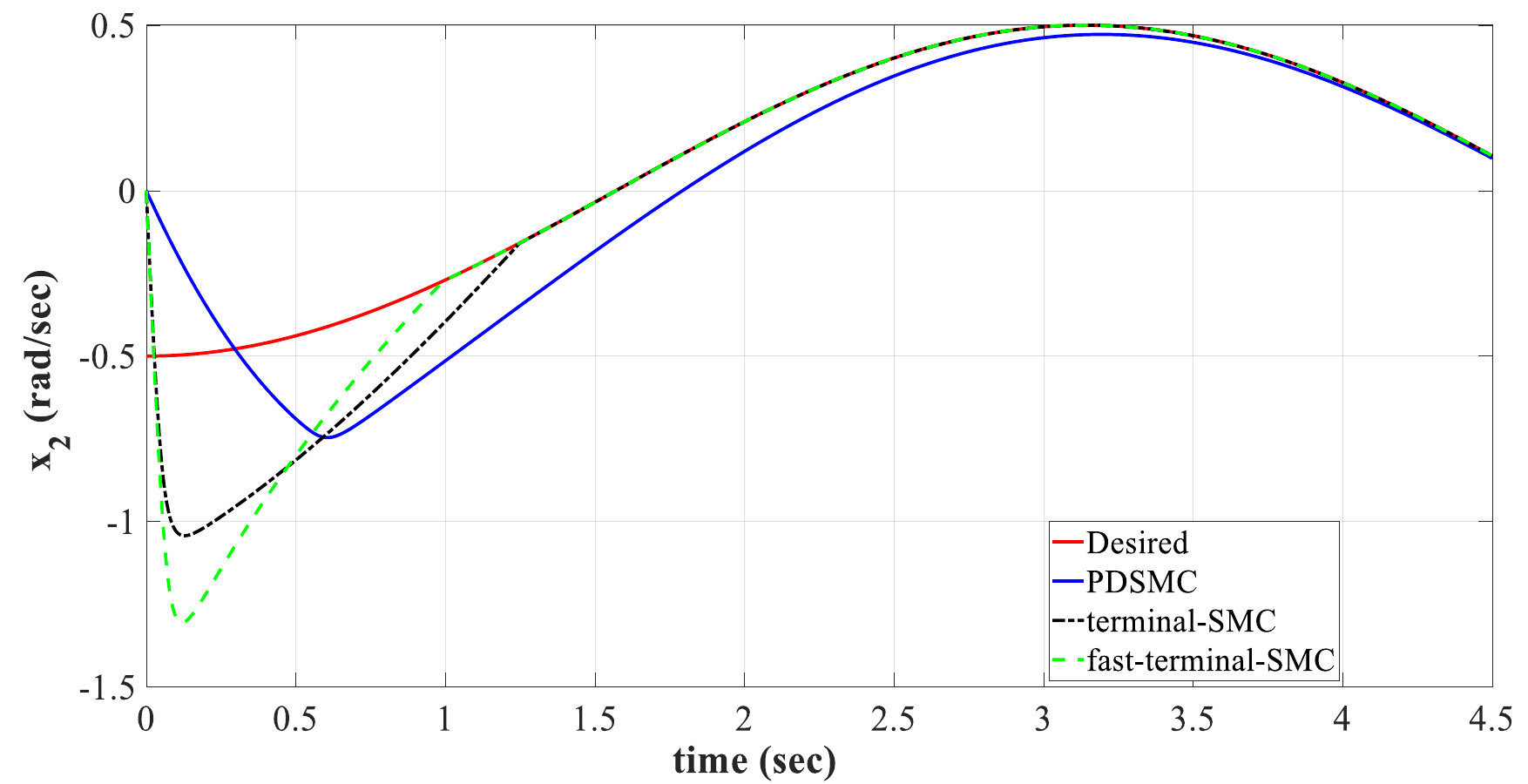}
	\caption{Angular velocity with conventional PD SMC, the terminal SMC, and the fast terminal SMC with hyperbolic tangent function}
	\label{}
\end{figure}
\begin{figure}
	\centering
	\includegraphics[width=1\linewidth]{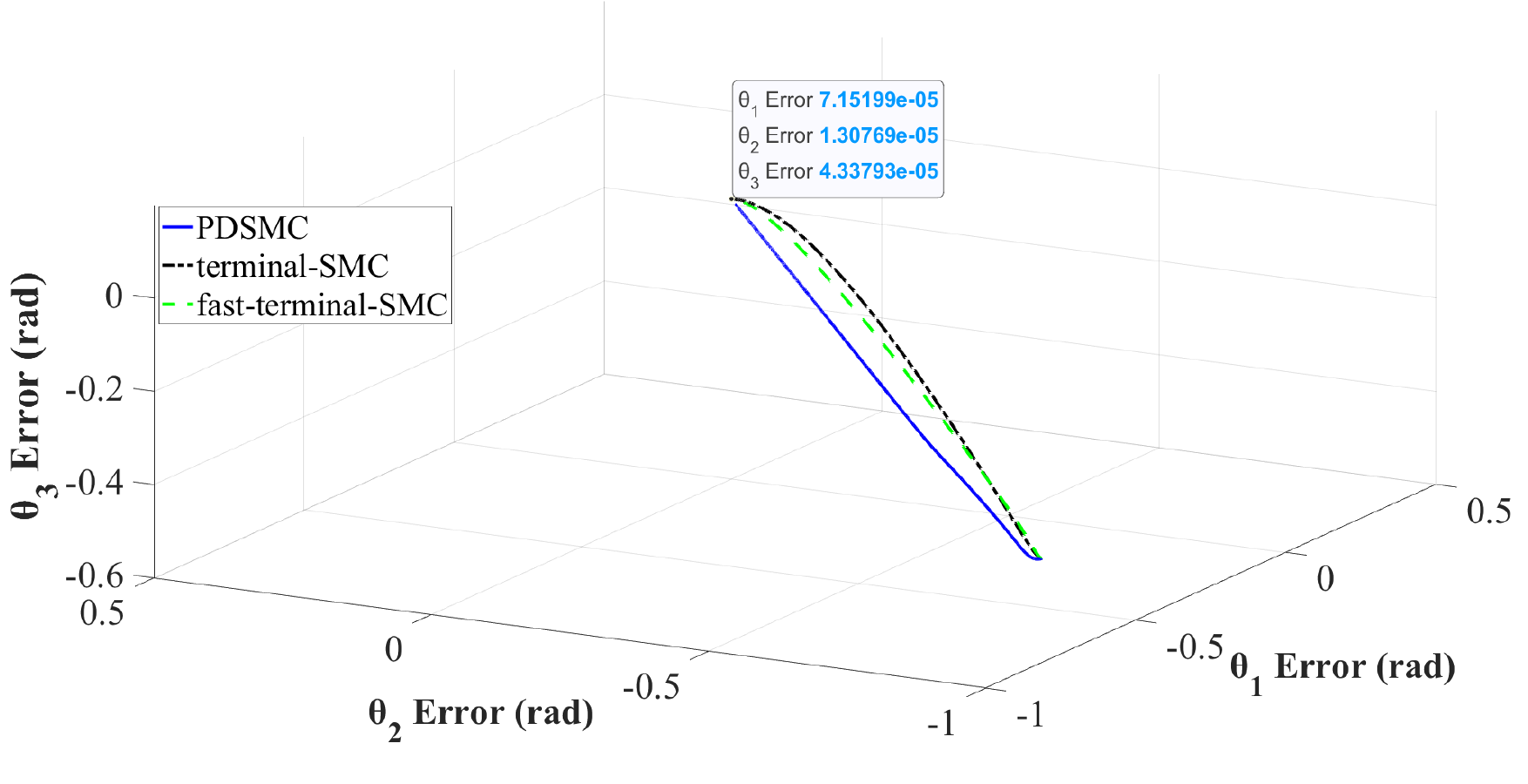}
	\caption{Position tracking error with conventional PD SMC, the terminal SMC, and the fast terminal SMC with hyperbolic tangent function}
	\label{}
\end{figure}
\begin{figure}
	\centering
	\includegraphics[width=1\linewidth]{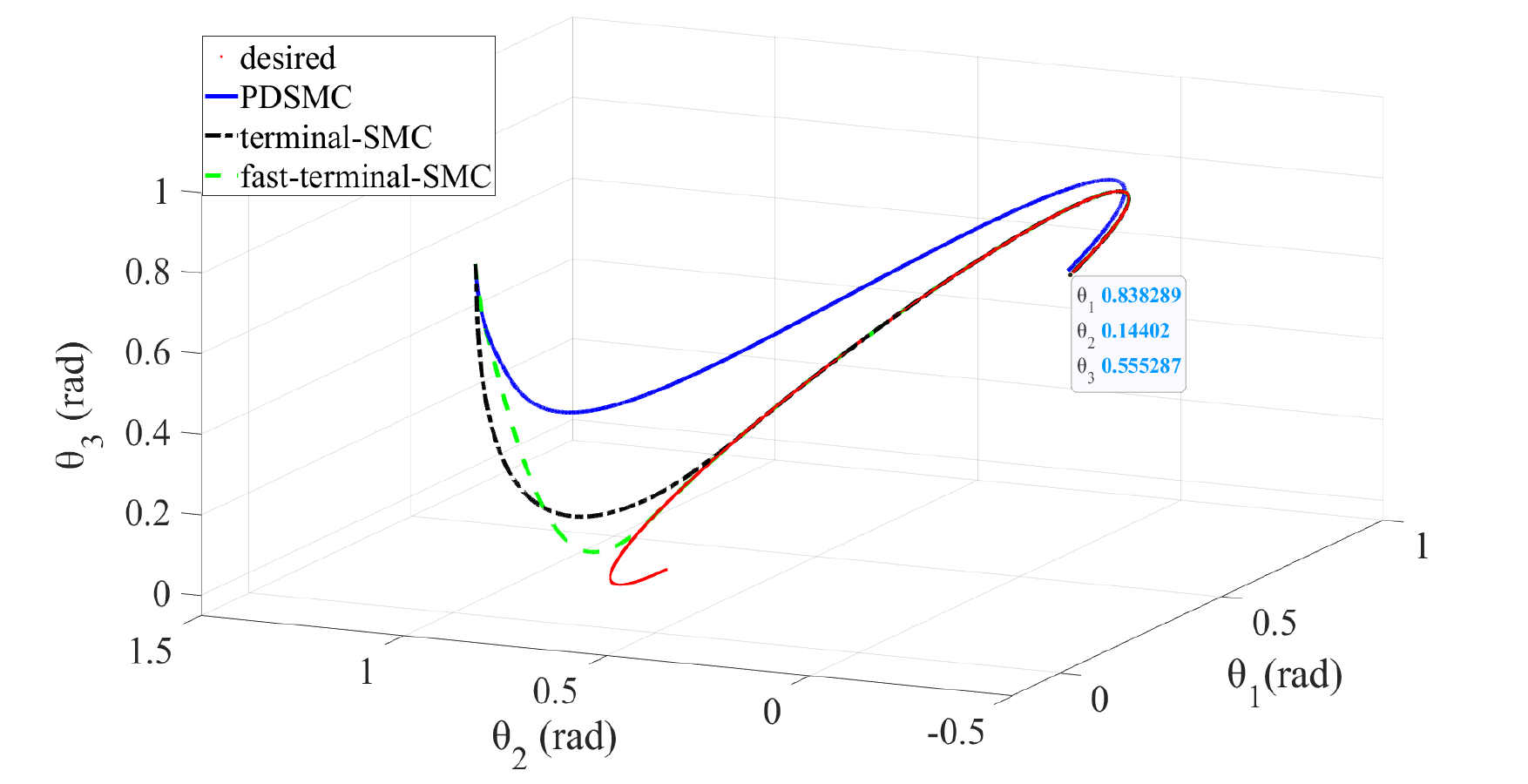}
	\caption{Position tracking with conventional PD SMC, the terminal SMC, and the fast terminal SMC with hyperbolic tangent function}
	\label{}
\end{figure}

\subsection{Fast terminal sliding mode control in presence of external disturbances}

To analyze the control resistance of the second order sliding mode, a normally distributed random noise, which is considered an external disturbance, is applied to the arms’ positions, to simulate the unwanted affects which the environment of these robots can have, usually in industrial fields.
Similar to the previous section, the FTSMC has shorter reaching times and high accuracy even in the presence of external disturbances, as it is shown in figures 16-18. 
Angular velocity however is somehow affected by the disturbance, however the FTSMC suppresses the affect much more compared to the other controllers, shown in figure 19.
Trajectory tracking errors shown in figures 20-22, confirms that the higher accuracy of FTSMC and steady-state zero state error.

\begin{figure}[H]
	\centering
	\includegraphics[width=1\linewidth]{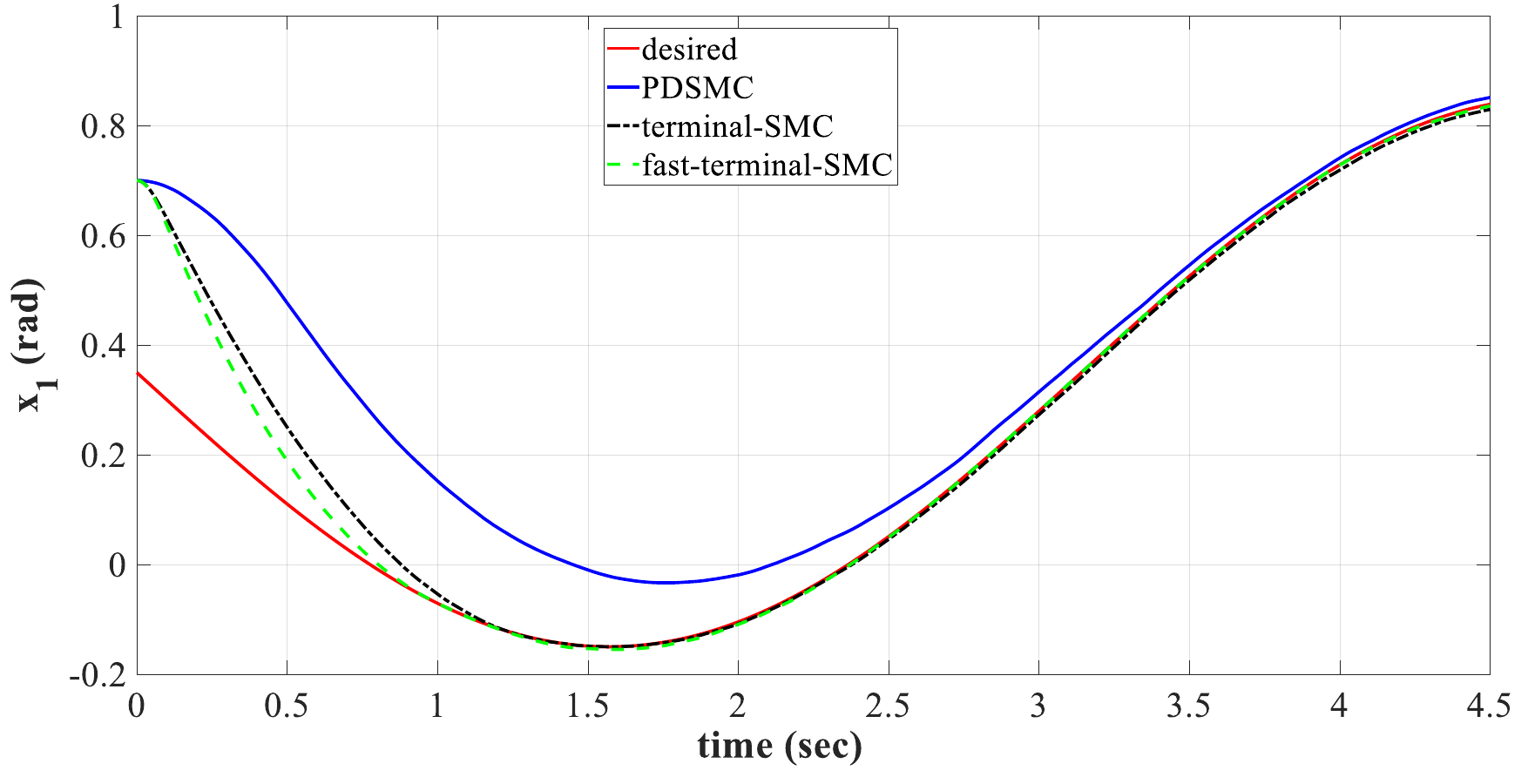}
	\caption{Tracking of the first arm of the robot with conventional PD SMC, the terminal SMC, and the fast terminal SMC in the presence of noise}
	\label{}
\end{figure}
\begin{figure}[H]
	\centering
	\includegraphics[width=1\linewidth]{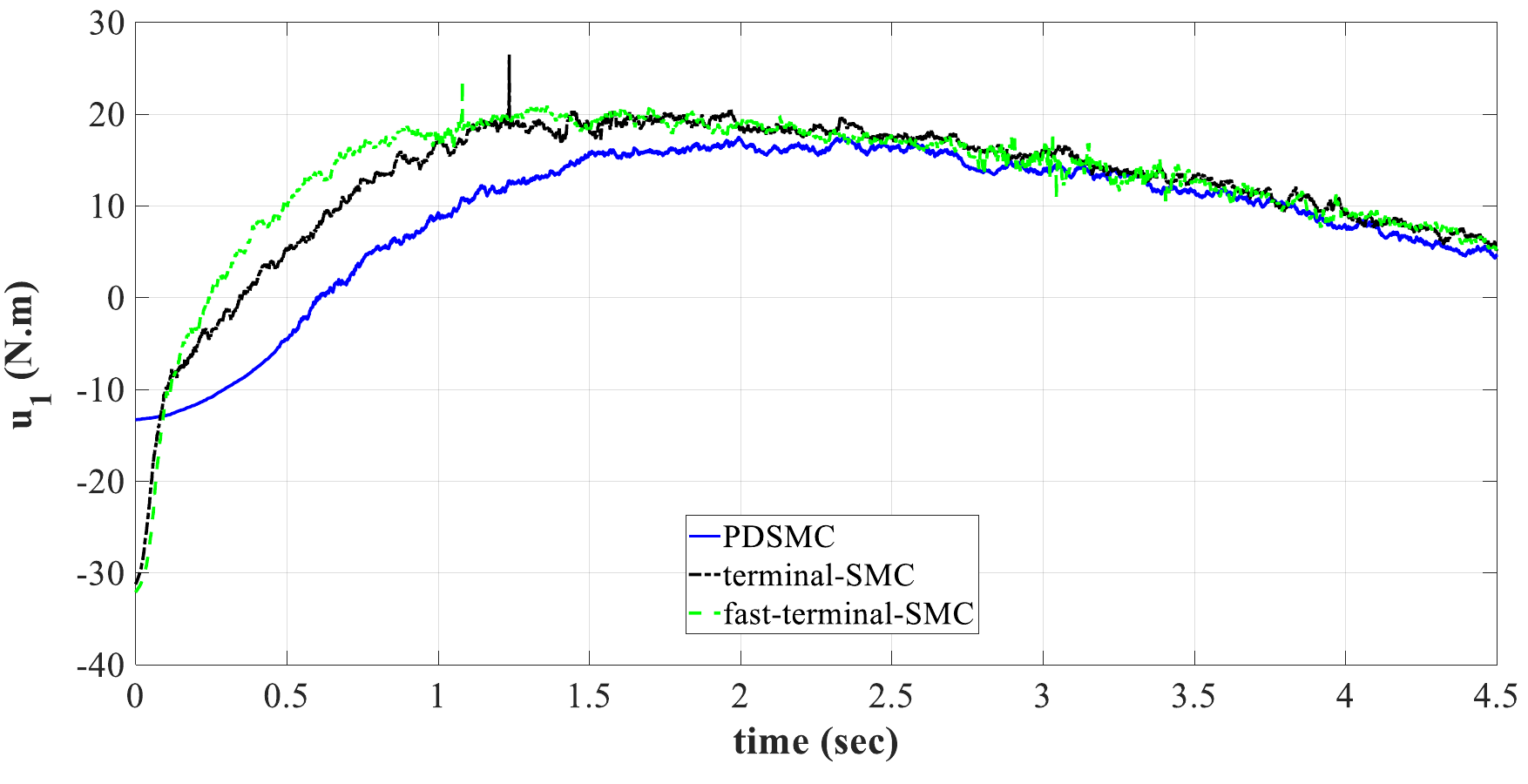}
	\caption{Control effort for the movement of the first arm of the robot with conventional PD SMC, the terminal SMC, and the fast terminal SMC in the presence of noise}
	\label{}
\end{figure}
\begin{figure}[H]
	\centering
	\includegraphics[width=1\linewidth]{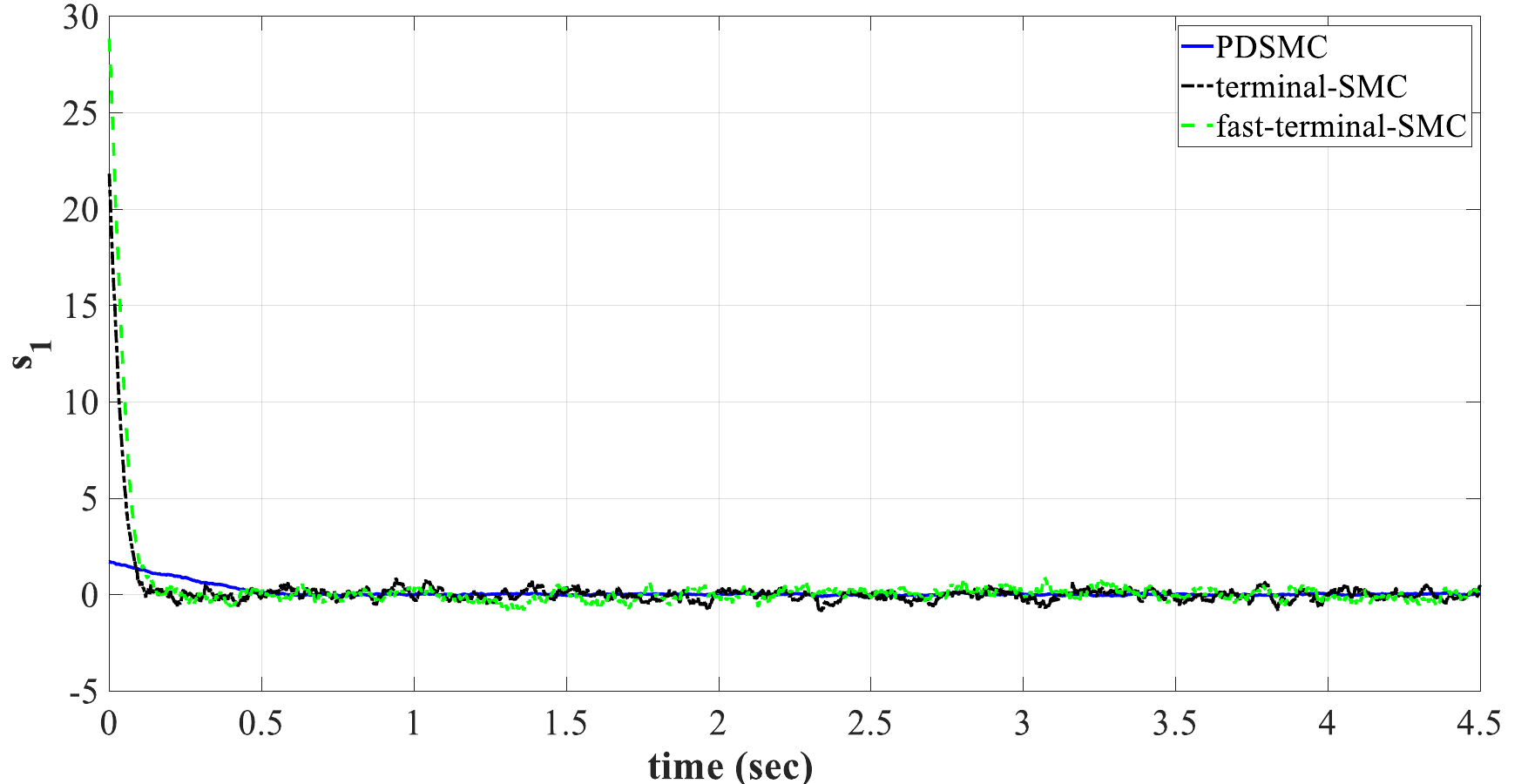}
	\caption{Reaching time to sliding surface with conventional PD SMC, the terminal SMC, and the fast terminal SMC in the presence of noise}
	\label{}
\end{figure}
\begin{figure}[H]
	\centering
	\includegraphics[width=1\linewidth]{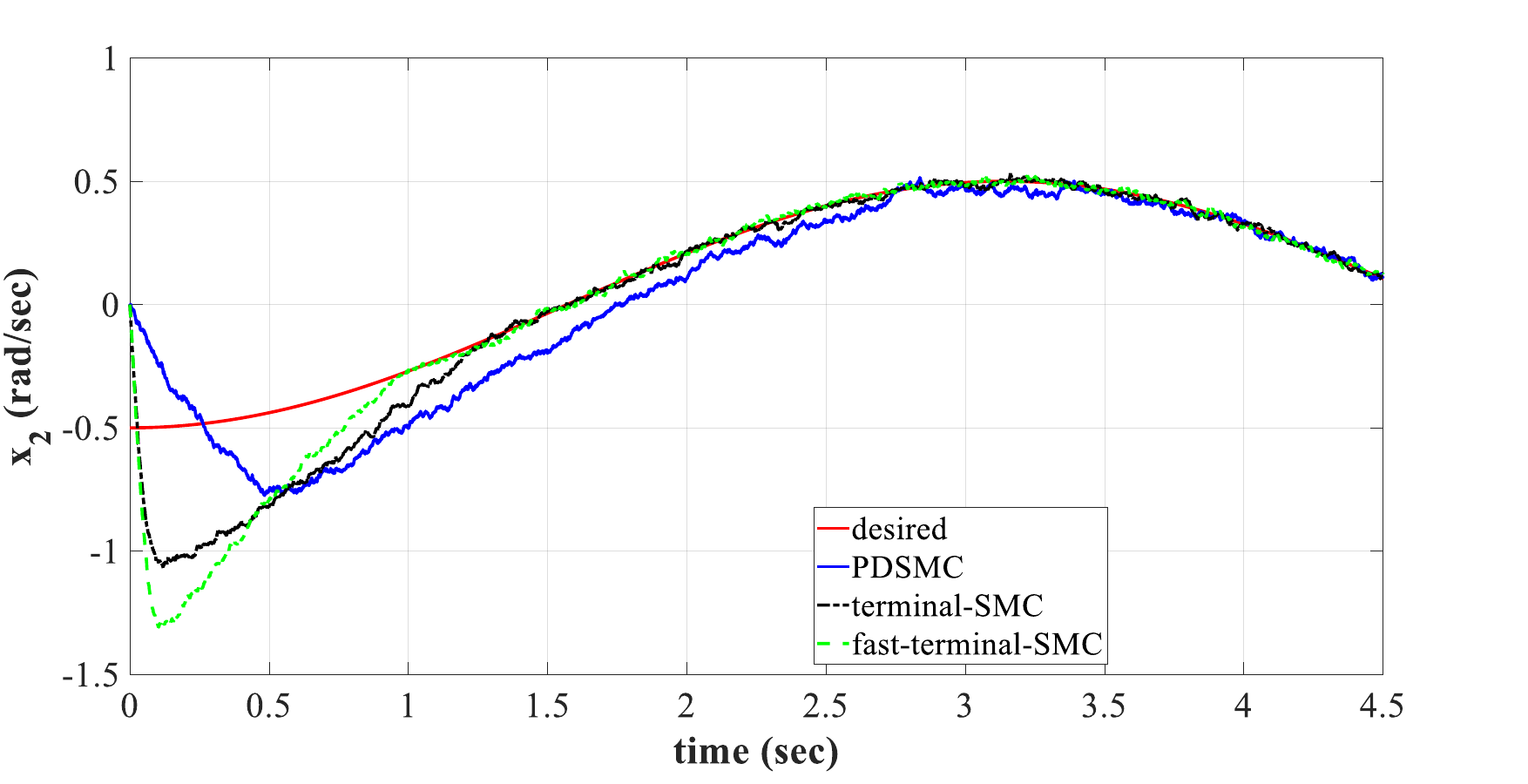}
	\caption{Angular velocity with conventional PD SMC, the terminal SMC, and the fast terminal SMC in the presence of noise}
	\label{}
\end{figure}[H]
\begin{figure}
	\centering
	\includegraphics[width=1\linewidth]{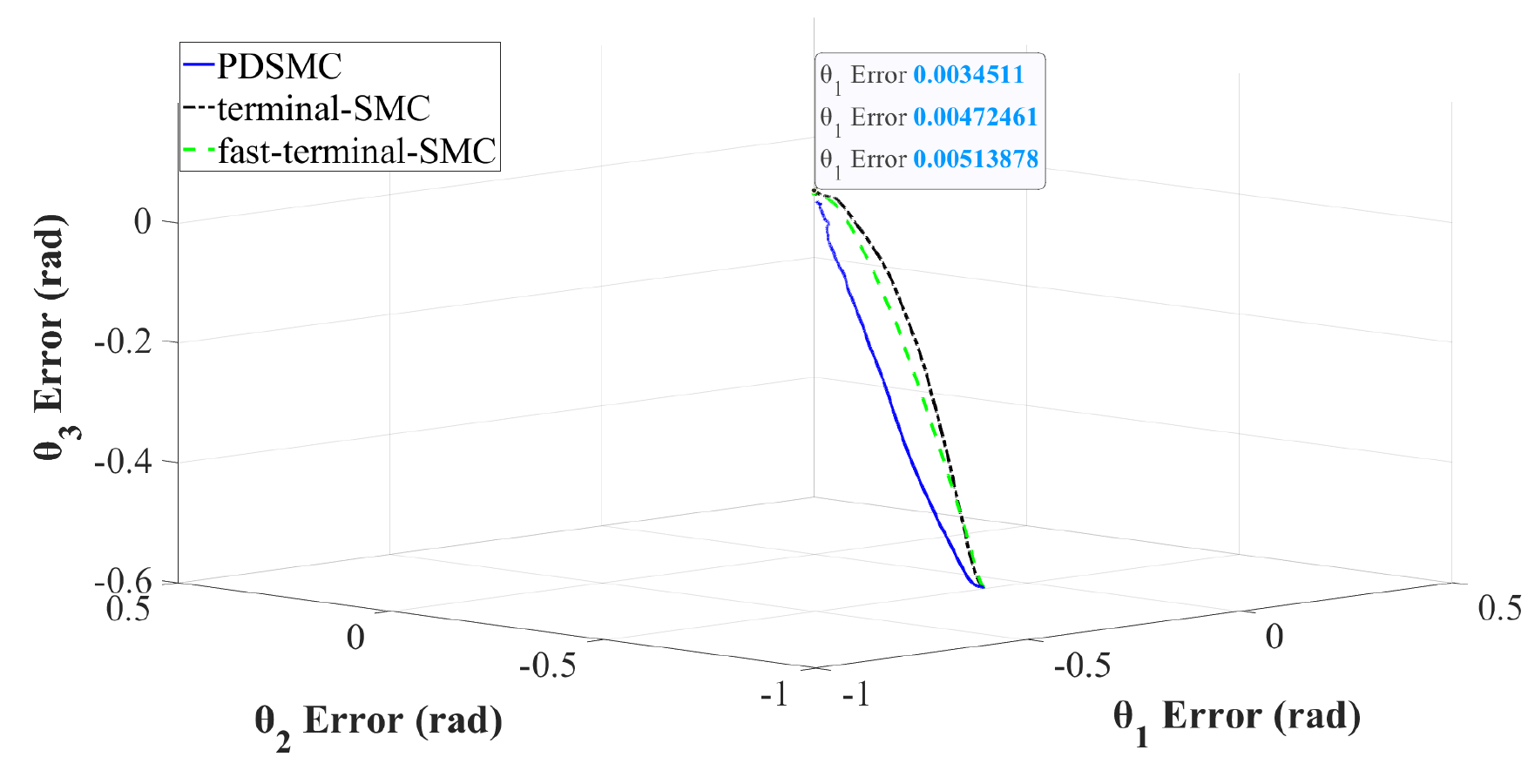}
	\caption{Position tracking error with conventional PD SMC, the terminal SMC, and the fast terminal SMC in the presence of noise}
	\label{}
\end{figure}[H]
\begin{figure}
	\centering
	\includegraphics[width=1\linewidth]{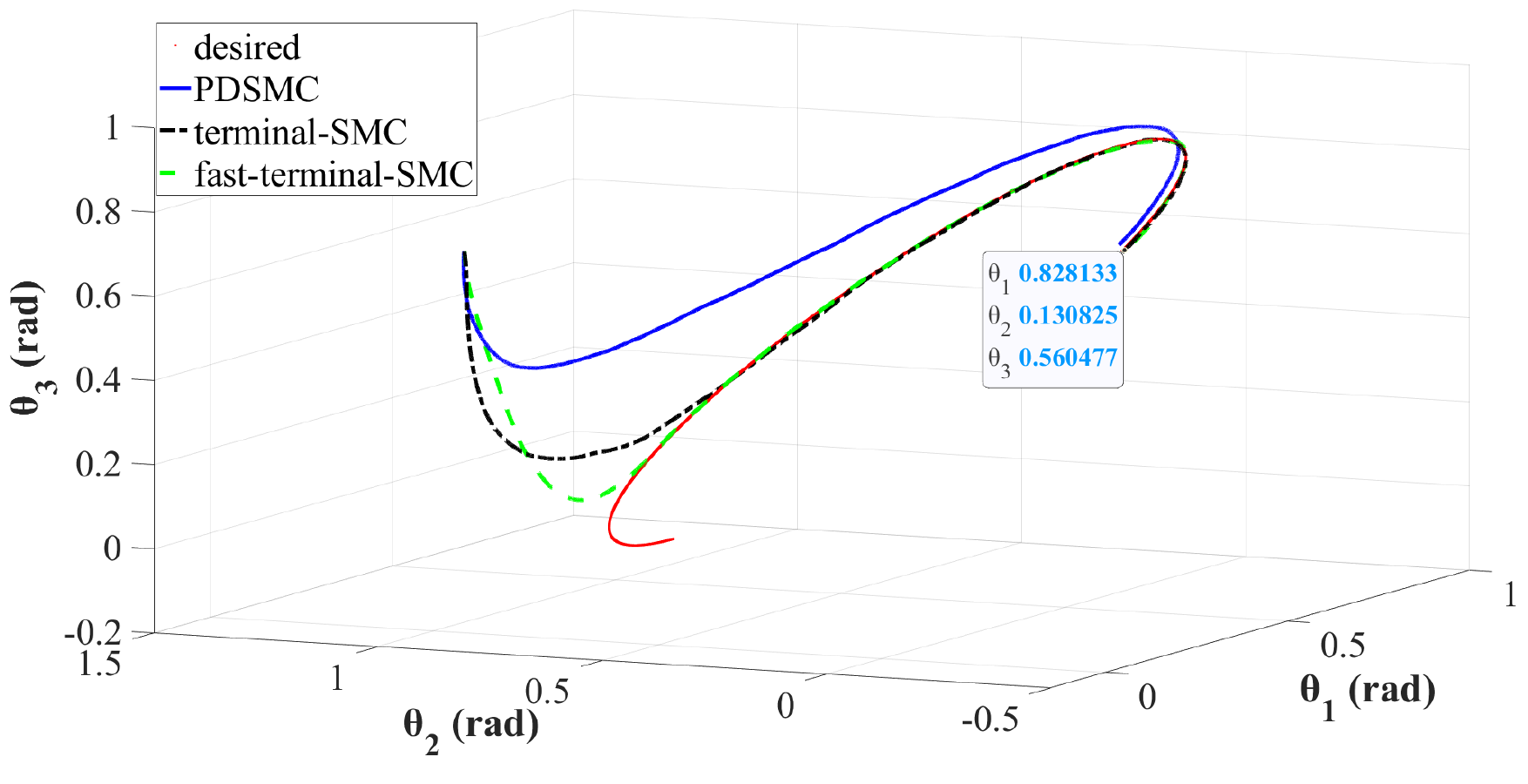}
	\caption{Position tracking with conventional PD SMC, the terminal SMC, and the fast terminal SMC in the presence of noise}
	\label{}
\end{figure}

\section{Conclusion} \label{{sec:conclusion}}
\begin{figure}[H]
	\centering
	\includegraphics[width=1\linewidth]{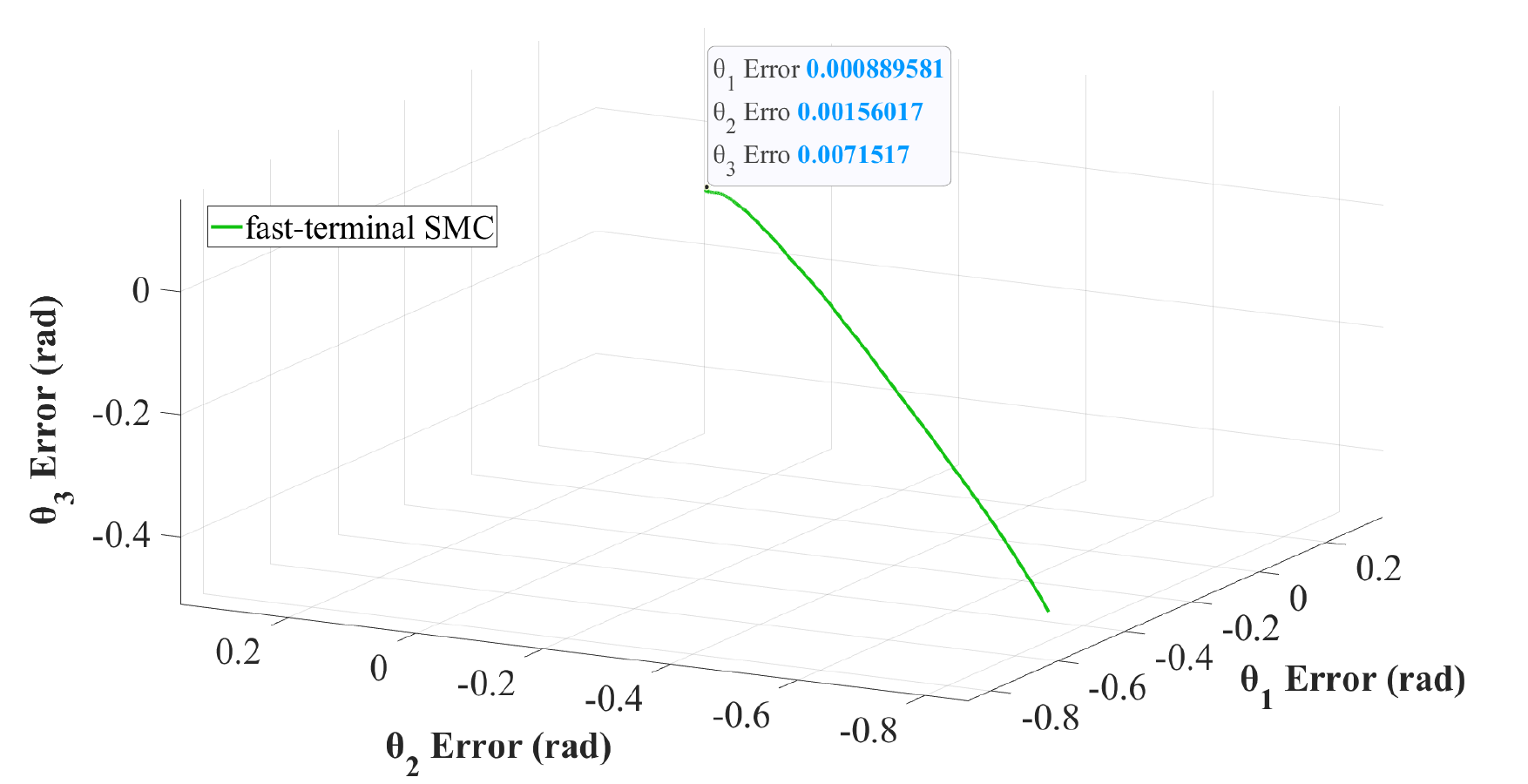}
	\caption{Steady-state error in position tracking of the proposed method}
	\label{}
\end{figure}

This paper focuses on proposing an FTSMC for the robotic arm. The state-space equations extracted from the dynamic equation of the robot represent both positions and velocities of each arm in 3-DOF. Simulation results show that FTSMC can track the desired position of each arm with the tracking error converging to zero ($e_1=-3.8\times10-7, e_2=5.58\times10-7, e_3=1.17\times10-6$)  in finite time. A hyperbolic tangent switching control law resolves the chattering phenomenon well and smooths control and output signals, as shown in Figures 11 and 13. A Lyapunov function is also proposed to prove the stable closed-loop performance and guarantee finite-time convergence to the designed sliding surface. The comparison between FTSMCT, TSCM, and conventional PDSMC closed-loop responses confirms the better performance of the designed controller in all performance characteristics than two other controllers, including shorter reaching time to sliding surface and higher tracking accuracy for all arms. Furthermore, in the presence of uncertainties, the FSTMC suppresses the effects of disturbances well without compromising the tracking accuracy, as displayed in figure 22. However, figure 18 shows the disturbance has some effect on the angular velocities of each arm, which equation (19) explains that the derivation of angular position amplifies the noise effects.
\begin{equation} 
\frac{d}{dt}x(t)\quad \underleftrightarrow{F} \quad j\omega X(j\omega)
\end{equation}
Although Figure 19 shows the FTSMC suppresses the noise effect for arms’ velocities, the TSMC and conventional PDSMC cannot wholly remove the unfavored effects of uncertainties. Finally, it can be concluded that the FTSMC has superior performance compared to conventional PDSMC and TSMC and meets all control goals determined for the robotic arm studied in this paper.\\
For future work, it will be interesting to investigate the effects of the parameter uncertainties and sensor faults in the design. Also, a learning-based FTSMC can be developed to consider the real nonlinear data of the robot instead of the equivalent dynamic relationship.
\bibliographystyle{unsrt}
\bibliography{ref.bib} 

\end{document}